\documentclass[%
reprint,
 amsmath,amssymb,
 aps,
]{revtex4-2}

\usepackage{graphicx}% Include figure files
\usepackage{dcolumn}% Align table columns on decimal point
\usepackage{bm}% bold math

\newcommand{\rem}[1]{}

\begin{document}

%\title{Mean-field theory for non-linear opinion dynamics with confirmation bias}
\title{Mean-field analysis for cognitively-grounded opinion dynamics with confirmation bias}

\author{Sven Banisch and Joris Wessels}
\affiliation{
Karlsruhe Institute of Technology, Institute for Technology Futures (KIT-ITZ)
}

%\title{Mean-field theory for non-linear cognitively grounded opinion dynamics with confirmation bias}

%\shorttitle{Influence response function}

% Use letters for affiliations, numbers to show equal authorship (if applicable) and to indicate the corresponding author

%\author[1,\Letter]{KIT Group}
%\affil[1]{Karlsruhe Institute of Technology, Institute for Technology Futures (KIT-ITZ)}

%

\begin{abstract}
Understanding how individuals’ beliefs and attitudes evolve within a population is crucial for explaining social phenomena such as polarization and consensus formation. We explore a persuasive arguments model incorporating confirmation bias, where individuals preferentially accept information aligning with their existing beliefs. By employing a mean-field approach, widely used in statistical physics, we simplify complex processes of argument exchange within the population. Our analysis proceeds by projecting the model onto continuous opinion dynamics and further reducing it through mean-field reasoning. The findings highlight the robustness of mean-field predictions and their compatibility with agent-based simulations, capturing the transition from consensus to polarization induced by confirmation bias.
\end{abstract}

\maketitle

\section{Introduction}

Understanding how individuals’ beliefs and attitudes evolve within a population is crucial for explaining social phenomena such as polarization, consensus formation, and the spread of misinformation. Opinion dynamics models have been instrumental in providing insights into these processes by simulating the interactions and influences among individuals \cite{axelrod1997dissemination,deffuant2000mixing,hegselmann2002opinion,Flache2017models,Banisch2019opinion}. Our research explores a persuasive arguments model \cite{Banisch2023biased} where agents exchange pro and con arguments, thereby shaping their opinions through social interactions. %\cite{Maes2013differentiation,Banisch2021argument,Taillandier2021introducing}.

A key element of the model is the incorporation of confirmation bias, a cognitive mechanism where individuals preferentially accept information that aligns with their existing beliefs while discounting contradictory evidence. This bias is known to play a significant role in real-world opinion dynamics \cite{lord1979biased,nickerson1998confirmation,knobloch2015political,ling2020confirmation}, leading to the reinforcement of existing beliefs and the potential for increased polarization within a population \cite{Dandekar2013biased,Lorenz2021individual,Banisch2023biased,lefebvre2024roots}.

To analyze the complex socio-cognitive interactions within the population, we employ a mean-field approach. 
Mean-field theory, widely used in physics to simplify complex systems, has been applied in various domains such as 
%condensed matter physics \cite{pitaevskii2016bose}, 
epidemics \cite{kermack1927contribution} and neural networks \cite{amit1985spin}. The use of mean-field theory in social dynamics has been extensively reviewed by Castellano et al. \cite{castellano2009statistical}, highlighting its effectiveness in capturing the macroscopic behavior of social systems from microscopic interactions. 
%By drawing parallels to these physical systems, we simplify our opinion dynamics model by considering the average effect of interactions rather than tracking each interaction individually.

Persuasive argument models \cite{Maes2013differentiation,Banisch2021argument,Taillandier2021introducing,Banisch2023biased} explicitly model a cognitive layer of arguments. Therefore, model reduction proceeds in two main steps. First, we project the original model \cite{Banisch2023biased} onto the space of continuous opinion dynamics \cite{Flache2017models,Lorenz2007continuous}, deriving an influence response function (IRF, \cite{lopez2008social,reitenbach2024coupled}) that governs the expected opinion change. This projection enables a comparison between the reduced and the original model through simulations. Second, we further reduce the model by a mean-field reasoning, dividing the population into two compartments. Using dynamical systems tools, we provide a comprehensive examination of the critical points and transitions in this idealized compartment model. The study demonstrates that mean-field treatment can effectively simplify the analysis of cognitively-grounded opinion dynamics, capturing essential features such as critical points and phase transitions to a high degree of accuracy.

% it also enables systematic model comparisons

\section{Persuasive arguments model}

The first agent-based model (ABM) drawing on persuasive arguments theory \cite{burnstein1977persuasive,hinsz1984persuasive} was introduced by Mäs and Flache in 2013 \cite{Maes2013differentiation}. They showed that opinion polarization can emerge if agents preferentially exchange arguments with similar others (homophily), leading to separated argument pools \cite{sunstein2002law}. Subsequent studies have analyzed this consensus-polarization transition for various settings \cite{Feliciani2020persuasion,Banisch2021argument}, including confirmation bias \cite{Taillandier2021introducing,Banisch2023biased,banisch2024validating}. 

In the model analyzed in this letter \cite{Banisch2023biased} agents hold an opinion on a scale from $-M$ to $+M$. $M$ denotes the number of pro and con arguments that agents can belief to be either true or false ($b_{ik} \in \{0,1\}$). The opinion of an agent $i$ is given by the number of pro versus con arguments. Assuming that the first $M$ elements of the argument vectors $b_i$ are pro and the remaining $M$ elements are con arguments, this can be written as
\begin{equation}
    o_i = \sum_{k = 1}^{M} b_{ik} - \sum_{k = M+1}^{2M} b_{ik}.
    \label{eq:opinioncomputationfori}
\end{equation}
In interaction, agents do not influence their opinions directly (as opposed to most other opinion models), but exchange arguments. That is, at each time step, two agents $i$ and $j$ are chosen at random. Agent $j$ is considered as a sender and articulates a random argument $b_{jk}$ from their current string to agent $i$. 

In order to integrate confirmation bias, \cite{Banisch2023biased} assume that the probability that $i$ accepts the new argument depends on $i$'s current opinion with 
\begin{equation}
    p_{\beta}(o_i,v) = \frac{1}{1+e^{-\beta o_i v}}
    \label{eq:argadoption}
\end{equation}
where $v \in \{-1,1\}$ encodes whether $b_{jk}$ contributes to a positive or negative stance. That is, if the argument confirms the current opinion, the acceptance probability is greater than $1/2$, and smaller if the argument challenges $i$'s current opinion. The strength of the confirmation bias is governed by $\beta$. Notice that positive (negative) contributions can be obtained either by a new pro (con) argument or by dropping an existing con (pro) argument. Finally, if agent $i$ accepts the argument ($b_{ik} = b_{jk}$), the opinion $o_i$ is updated by Eq.\eqref{eq:opinioncomputationfori}.

\section{Projection onto continuous opinion dynamics}

Most models of continuous opinion dynamics assume that opinions range from -1 to 1 \cite{Lorenz2007continuous,Flache2017models,Lorenz2021individual}. They further assume that in interaction the opinion of an agent $i$ is influenced by the opinion of $j$ by an IRF governing $i$'s opinion change such that $\Delta o_i = f(o_i,o_j)$. 
In the sequel, we will derive a formulation of the PAT model on a continuous opinion scale by computing the expected opinion change $\Delta o_i = f(o_i,o_j) := \mathrm{E}[\Delta o|o_i,o_j]$ which only depends on the opinions (and not on the beliefs) of the two agents.

%In order to be able to compare the PAT model to these other models  it will be useful to 

To achieve that, we first normalize the opinion computation. For further convenience, we introduce the bracket notation $\langle \rangle_p$ and $\langle \rangle_c$ as the normalized sum over the pro and the con part of the argument strings
\begin{equation}
     \langle b_i \rangle_p = \frac{1}{M}\sum_{k=1}^M b_{ik} \ \text{ and } \
    \langle b_i \rangle_c = \frac{1}{M}\sum_{k=M+1}^{2M} b_{ik}.   
\end{equation}
The rescaled opinion is then given by $o_i = \langle b_i \rangle_p - \langle b_i \rangle_c$ \footnote{Note that this rescaling has to be carried over to Equation (\ref{eq:argadoption}) by adjusting the parameter that governs the strength of biased processing $\beta \rightarrow M \beta$.
Hence, while $\beta = 0.5$ has been identified as a critical value in a model with $M = 4$ pro and con arguments \citep{Banisch2023biased}, this corresponds to $\beta = 2$ after rescaling.}.

Next, we derive the expected opinion change $\mathrm{E}[\Delta o|o_i,o_j]$ once two agents with opinions $o_i$ and $o_j$ have been chosen. 
We first consider the case that $j$ supplies $i$ with a new argument in favor of the issue. Recall that favorable information may correspond to a new pro argument ($b_{jk} = 1, b_{ik} = 0$ for $1 \leq k \leq M$) or to a counter argument that is dismissed ($b_{jk} = 0, b_{ik} = 1$ for $M+1 \leq k \leq 2M$). Hence, the probability of new positive information is
\begin{equation}
    P[ j \xrightarrow[]{+} i] =
    \frac{1}{2} \Big( \langle b_j \rangle_p - \langle b_i b_j \rangle_p + \langle b_i \rangle_c - \langle b_i b_j \rangle_c
    \Big),
    \label{eq:newpositive02}
\end{equation}
where the two terms $\langle b_i b_j \rangle_p$ and $\langle b_i b_j \rangle_c$ are the normalized sums over element-wise products of $i$'s and $j$'s beliefs. Hence, these operators correspond to the normalized number of shared pro and con arguments respectively.
Equivalently, the probability that $j$ supplies $i$ with new information supporting a negative opinion is given by
\begin{equation}
    P[ j \xrightarrow[]{-} i] =
    \frac{1}{2} \Big( \langle b_i \rangle_p - \langle b_i b_j \rangle_p + \langle b_j \rangle_c - \langle b_i b_j \rangle_c
    \Big).
    \label{eq:newnegative02}
\end{equation}

To compute the expected opinion change note that $o_i$ will increase by $1/M$ if $i$ accepts a new positive information and decrease by $1/M$ after accepting negative information.
With the acceptance probability in Eq. \eqref{eq:argadoption},
the expected opinion change is given by
\begin{equation}
    \mathrm{E}[\Delta o_i|b_i,b_j] = \frac{1}{M} \Big( p_{\beta}(o_i,+) P[ j \xrightarrow[]{+} i] - p_{\beta}(o_i,-) P[ j \xrightarrow[]{-} i] \Big)
    \label{eq:sophisticatedsolution}
\end{equation}
In order to arrive at an IRF that depends only on opinions, we exploit the fact that $p_{\beta}(o_i,+) - p_{\beta}(o_i,-) = \tanh(\frac{o_i \beta}{2})$ and the covariance relations $\sigma_p(b_i,b_j) = \langle b_i b_j\rangle_p - \langle b_i\rangle_p \langle b_j\rangle_p$ and $\sigma_c(b_i,b_j) = \langle b_i b_j\rangle_c - \langle b_i\rangle_c \langle b_j\rangle_c$ 
(see Appendix \ref{sec:SI:A:sophi}). 
With $o_i = \langle b_i \rangle_p - \langle b_i \rangle_c$ and $o_j = \langle b_j \rangle_p - \langle b_j \rangle_c$ we arrive at 
\begin{equation}
    \mathrm{E}[\Delta o_i|b_i,b_j] = \frac{1}{4M} \Big[ o_j - o_i + \tanh\Big(\frac{o_i \beta}{2}\Big) \Big(1 - o_i o_j + \hat{y}(b_i,b_j)\Big)\Big] 
\end{equation}
where only the term $\hat{y}(b_i,b_j)$ depends on the belief strings. 
It hence remains to show that the means of $\hat{y}(b_i,b_j)$ over all belief strings $b_i$ and $b_j$ that map onto the same opinions $o_i$ and $o_j$ is zero for any opinion pair. For the model with $M = 4$, we explicitly evaluated $\hat{y}(b_i,b_j)$ for all possible combination of belief strings (there are a total of $2^{2M}\times2^{2M}$ possible combinations) and show that this is the case (see Appendix \ref{sec:SI:A:inter}).

We hence obtain as an IRF derived from the PAT model
\begin{equation}
    \mathrm{E}[\Delta o_i|o_i,o_j] = \frac{1}{4M} \Big[ o_j - o_i + \tanh\Big(\frac{o_i \beta}{2}\Big) \Big(1 - o_i o_j \Big)\Big].
    \label{eq:naivesolution}
\end{equation}
Notice that this solution can be obtained also by a naive approach which does not take into account how many arguments $i$ and $j$ already share. See Appendix \ref{sec:SI:A:naive}. 

\section{Mean-field model}

The IRF derived with Eq. \eqref{eq:naivesolution} can be used to specify an ABM for non-linear continuous opinion dynamics. In the results section, we compare this IRF model to the original one and highlight some important differences. 
However, it turns out that an accurate mean field model can be obtained by a very simple ansatz, assuming that the population is split into two groups $A$ and $B$. Agents within the same group are homogeneous with respect to their opinion and interaction probabilities, but we may accommodate differences across the two groups.

We consider the most simple setting.
Let us assume that the two groups are of equal size. Let us further assume that an agent in group $A$ interacts with another agent in $A$ with probability $1-p$ and encounters an agent in $B$ with probability $p$. Vice versa for agents in $B$. Using Eq. \eqref{eq:naivesolution}, we obtain for the opinion change for groups $A$ and $B$
\begin{align}
      \Delta o_A &= \frac{\tanh \left(\frac{\beta o_A}{2}\right) \left(o_A^2 (p-1)-o_A o_B p+1\right)+p (o_B-o_A)}{4 M}\\ 
      \Delta o_B &= \frac{\tanh \left(\frac{\beta o_B}{2}\right) \left(o_B^2 (p-1)-o_A o_B p+1\right)+p (o_A-o_B)}{4 M}
\end{align}
It turns out that this MF model captures the dynamics of the original and the reduced ABM remarkably well, which we show in the next section.

\section{Results}

In this section, we demonstrate that the three models, each with decreasing complexity, exhibit similar behavior. The mean-field approach accurately captures the transition from consensus to polarization induced by increasing confirmation bias and provides detailed understanding concerning different initial conditions and their convergence under the ABM dynamics.

%\subsection{Stability of polarization and acceleration of consensus}
\subsection{Stability of polarization}

We first focus on the relation between the original PAT model and its reduced IRF-based version. The models behave very similar in important regards. For $\beta = 0$, they both converge to a moderate consensus state where all agents hold a neutral opinion. As soon as $\beta > 0$, an extreme consensus emerges with all agents holding opinion +1 or -1. As $\beta$ further increases, a bi-polar opinion distribution emerges with approximately one half of the population with positive and the other half with negative opinions (notice that $|o_i| < 1$). Hence, both ABMs capture the transition from consensus to polarization as the confirmation bias $\beta$ increases.

%However, in the original model, bi-polarization is a meta-stable state \cite{Banisch2023biased}. The longer the model is run, the more likely a polarized state collapses to extreme consensus. \cite{Banisch2023biased} have shown that the time during which polarization persists increases exponentially with $\beta$. In the reduced ABM the polarization state becomes stable. This is shown in Figure \ref{fig:panel_runtime} which shows the frequency of model realizations in a polarized state after a certain number of steps ($T = 800, 2000, 4000, 8000, 16000, 32000, 64000$). In the PAT model (left), more and more runs collapse into extreme consensus, shifting the transition towards higher levels of $\beta$. In the IF model (right), no such shift can be observed, meaning that once a bipolar attractor has been found, the system will stay there for very long times.

\begin{figure}[htbp]
    \centering
    \includegraphics[width=0.99\linewidth]{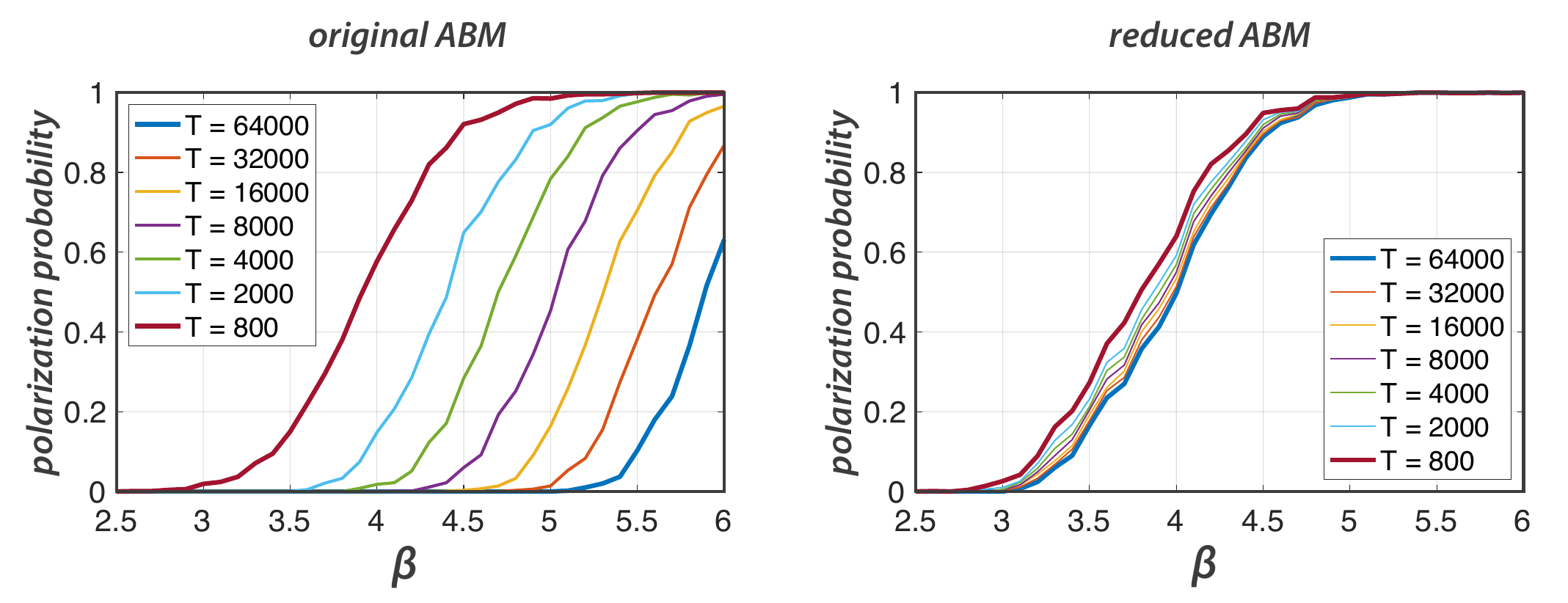}
    \caption{The meta-stable state of bi-polarization becomes stable in the reduced version of the model: Polarization rate after $T = 800, 2000, 4000, 8000, 16000, 32000, 64000$ steps for $N = 100$ and $M = 4$. Confirmation bias is sampled in between $2.5 \leq \beta \leq 6$ with step size $0.1$ and $10^3$ samples per point.}
    \label{fig:panel_runtime}
\end{figure}

In the original model, bipolarization is a metastable state, which is likely to collapse into extreme consensus over time. The duration for which polarization persists increases exponentially with the parameter $\beta$ \cite{Banisch2023biased}. In the reduced ABM, the polarization state becomes stable as shown in Figure \ref{fig:panel_runtime}. This figure illustrates the frequency of model realizations in a polarized state after varying numbers of steps ($T = 800, 2000, 4000, 8000, 16000, 32000, 64000$). In the PAT model (left), an increasing number of runs collapse into extreme consensus with rising $\beta$ levels, while the IRF model (right) shows no such shift: once a bipolar attractor is found, the system remains in that state for extended periods.%\footnote{Note that, in principle, the probability of leaving a bi-polarized state should still be non-zero??}

Note also that in the reduced model, the convergence to moderate consensus for $\beta = 0$ is significantly faster than in the more complex PAT model. Our model reduction approach hence enables quicker calculations and more efficient data processing, making it preferable in certain computational contexts.
%What is more, it also enables systematic comparison between PAT models to classical social influence models 

\subsection{Bifurcation analysis}

We now turn to the MF model. The MF model involves a tremendous reduction of complexity given both ABMs. It is based on the idea that two equally sized groups with homogeneous initial opinions engage in social influence dynamics on a complete graph ($p = 1/2$). We analyze this system in terms of its fixed points given by
\begin{align}
      \Delta o_A(o_A,o_B) & = 0\nonumber\\ 
      \Delta o_B(o_A,o_B) & = 0
      \label{eq:tosolve}
\end{align}
which we denote as $(o_A^*,o_B^*)$. For a given $\beta$, we can solve this system of equations with standard mathematical software (Mathematica). %and interactively explore the phase space of the model (see tools section). 

\begin{figure}[htbp]
    \centering
    \includegraphics[width=0.99\linewidth]{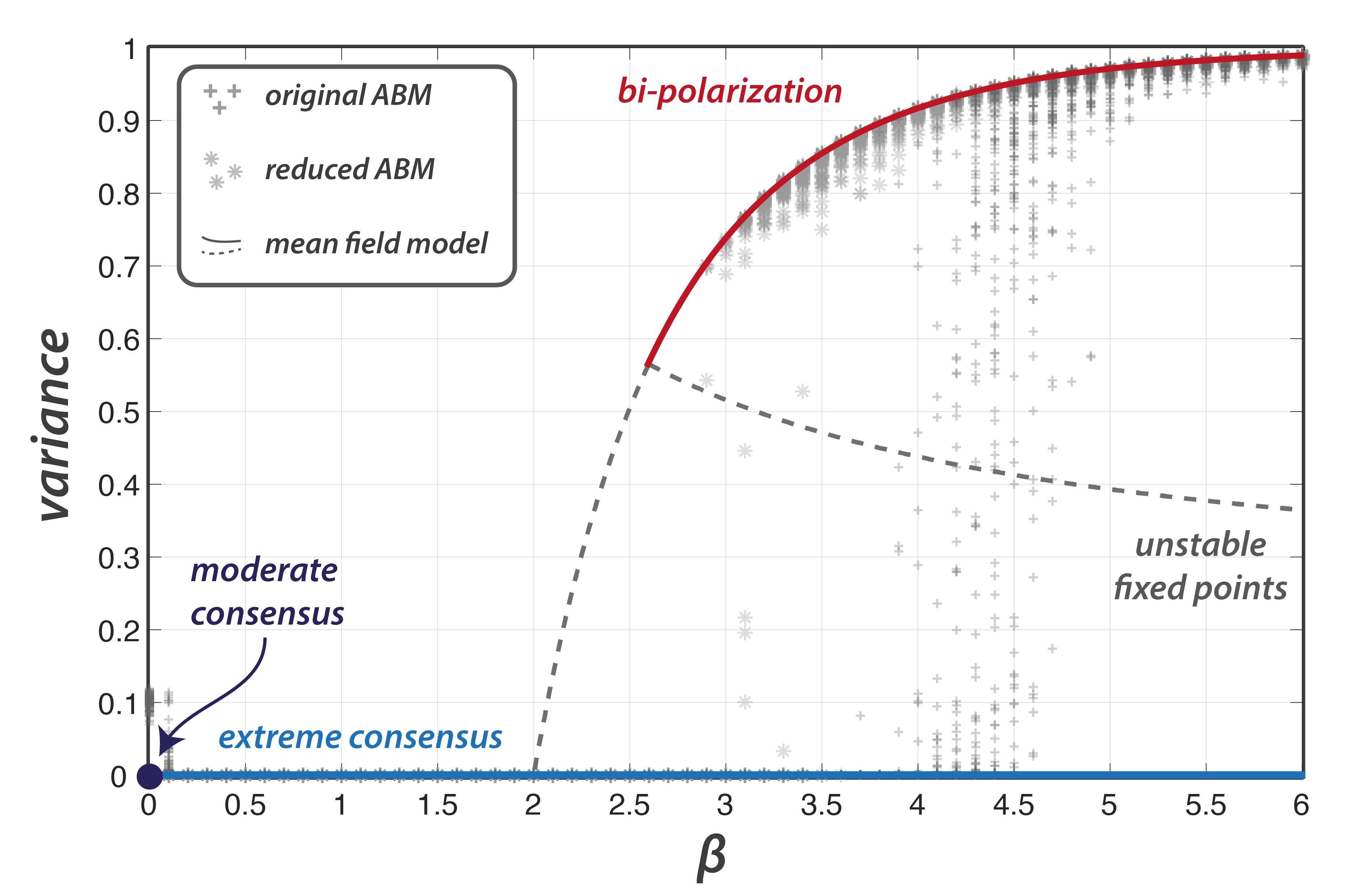}
    \caption{Basic bifurcation from consensus to polarization with increasing confirmation bias in terms of the variance of the final opinion profile. The MF solution (blue and red curves) are compared with repeated runs of both ABMs. Setting: $N = 1000, M = 4, T = 4000, \beta = [0,0.1,0.2,\ldots,6]$ and 100 runs per sample point. For each individual run, the mean variance over the last 400 steps is shown ($2 \times 6000$ points).}
    \label{fig:basicbifurcation}
\end{figure}

A global overview over the systems dynamics is provided in Fig. \ref{fig:basicbifurcation} with a bifurcation diagram for $\beta$. It is obtained by numerically solving Eq. \eqref{eq:tosolve} for $0 \leq \beta \leq 6$ (300 sample points). In order to compare to the ABM outcomes, we compute the variance over the fixed points $(o_A^*,o_B^*)$ distinguishing states of global consensus ($o_A^* = o_B^*$) from polarized states with $o_A^* \neq o_B^*$. As can be seen in Fig. \ref{fig:basicbifurcation}, the MF model accurately captures the basic transition from consensus to polarization with increasing confirmation bias. We can identify two critical values of confirmation bias at $\beta = 2$ and $\beta \approx 2.6$. Polarization only becomes stable for $\beta > 2.6$ (red curve). For values below that (but $\beta > 0$), the only stable solution is extreme consensus.

Fig. \ref{fig:basicbifurcation} also compares the MF solution to the ABMs in terms of the variance of the final opinion profile. For this purpose, the final states of the individual runs ($T = 4000$ steps) of a computational experiment are shown by the shaded stars and crosses for the PAT and the IRF model respectively. 
We observe a remarkable fit between MF theory and both ABMs. In the MF model, polarization and consensus are stable for $\beta > 2.6$. The reduced IRF model undergoes a sharp transition to polarization at $\beta \approx 3$. The final variance is accurately predicted by the corresponding bi-polarization fix points of the MF model. The original PAT model transitions from consensus to polarization in a parameter region of $4 < \beta < 5$. Notice, that the "final state" was measured on runs with $T = 4000$ steps, meaning that many realization were bi-polarized but have already collapsed before the end of the simulation (see previous section). But also for the more complex PAT model, final profiles are accurately captured by the MF approach. The two-group setting captures that the ABMs tend to produce nearly perfect bi-polarization into two equally sized groups, whose opinion strength ($|o|$) increases with confirmation bias.

\subsection{Initial conditions and finite size predictions}

One puzzle that remains from the previous analysis is that in the ABMs only polarization is observed for large $\beta$. In the MF model, the extreme consensus point remains a stable fixed point for all $\beta$, and this equilibrium is reached whenever $o_A$ and $o_B$ tend to the same side of the opinion spectrum. We show that this is an effect of heterogeneous initial conditions that project into a particular region of the MF space. 

\begin{figure}[htbp]
    \centering
    \includegraphics[width=0.99\linewidth]{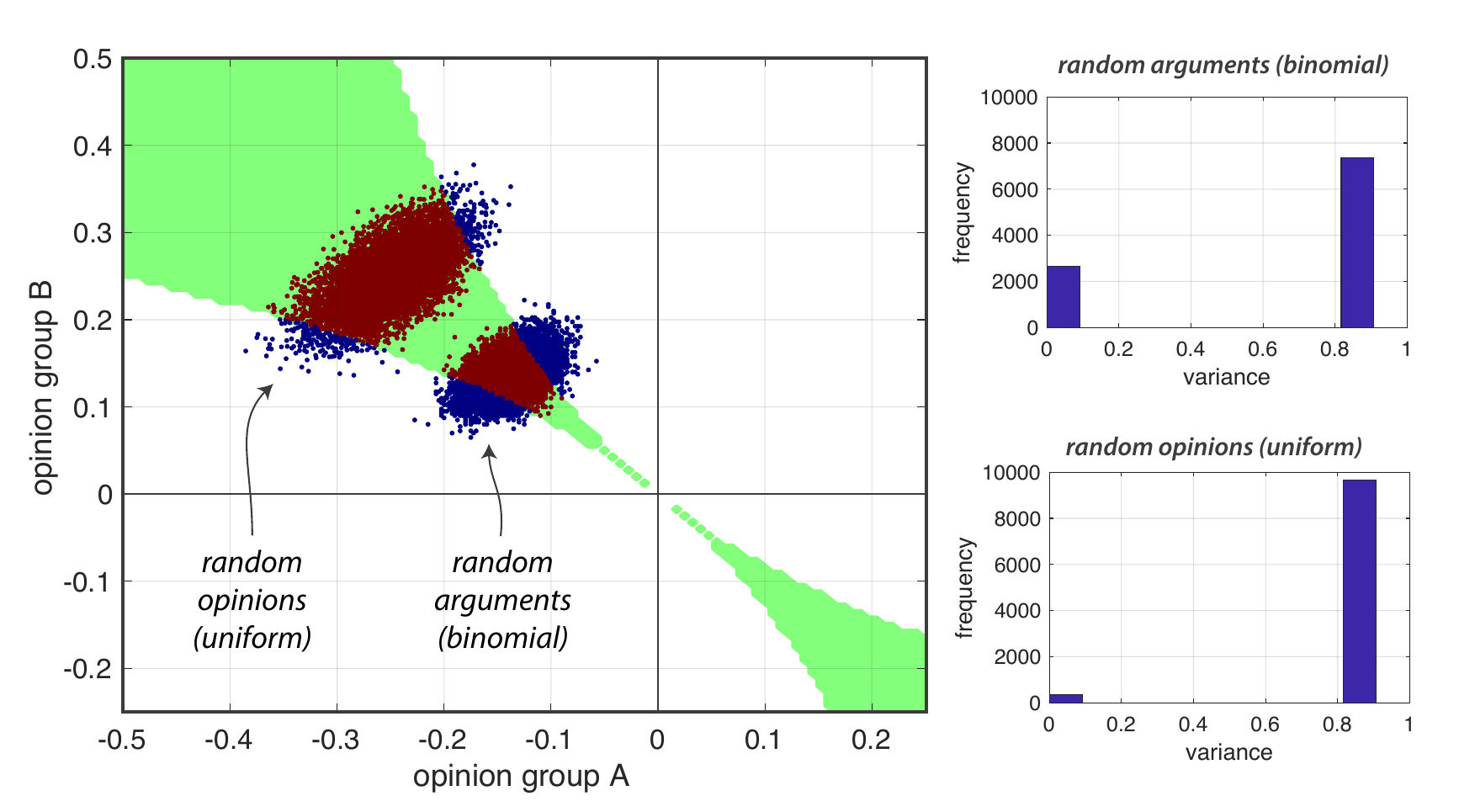}
    \caption{Projecting random initial condition used in the ABM with $N = 100$ into MF space. We compare random arguments (binomial) and random opinions (uniform). Both point clouds cut across the polarization basin of attraction for $\beta = 3.9$ (green shade) telling apart realizations that would converge to polarization or extreme consensus (histograms on the right).}
    \label{fig:ic:approach}
\end{figure}

In Fig. \ref{fig:ic:approach} we project random opinions for $N = 100$ agents into the MF space by dividing the agent population in two groups such that group A contains all agents with $o_i < 0$ and groups B all agents with $o_i > 0$ (\emph{splitting at zero}). We compare (i.) random arguments (leading to a binomial opinion distribution), and (ii.) uniformly at random ($o_i \in [-1,1]$). We draw $S = 10000$ samples capturing $10000$ repeated ABM runs (initial conditions).

Along with the projected initial conditions, we compute the basin of attraction for a given $\beta = 3.9$ (green region). We do that numerically by sampling regularly through the opinion space $(o_A,o_B)$ ($200 \times 200$ sample points, see Appendix \ref{sec:SI:D}). As can be seen, for $N = 100$, the projected initial conditions cut across the basin of attraction associated to the fixed point at the upper left where $o_A \approx -1$ and $o_B \approx 1$. Under the MF assumptions, this means that a certain part of the initial conditions will evolve to consensus (blue dots), and another part to polarization (red dots). This is shown by the two histograms on the right telling apart the two attractors.

\begin{figure*}[htbp]
    \centering
    \includegraphics[width=0.99\linewidth]{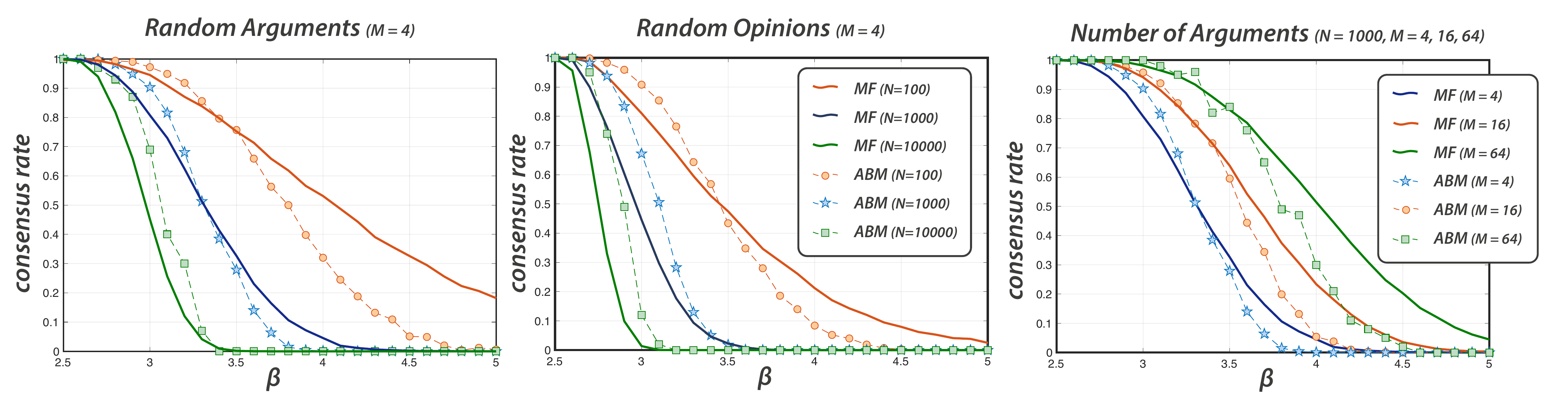}
    \caption{MF predictions of the consensus rate and comparison to simulations ($10^3$ runs per sample point). Left and middle: MF theory correctly predicts that the transition sharpens with increasing system size. Right: MF theory predicts the impact of increasing the number of arguments $M$ in the model. (See Appendix \ref{sec:B}.)}
    \label{fig:predictions}
\end{figure*}

    %Note that ABM $M = 64$ is based on 100 samples, all others on 1000. 
    
%\begin{figure}[htbp]
%    \centering
%    \includegraphics[width=0.99\linewidth]{figures/initialconditions/SummaryCRDifferentSplitting.pdf}
%    \caption{....}
%    \label{fig:ic:initialsumary}
%\end{figure}

This provides a prediction of the number of ABM realizations (respectively initialized) that will approach bi-polarization or extreme consensus.
The approach enables an extremely efficient method to study the impact heterogeneous initial conditions and basic model parameters such as different system sizes or the number of arguments. It also allows to assess the statistical impact of the number of samples $S$ in computational experiments. Fig. \ref{fig:predictions} demonstrates these capabilities by comparing MF predictions of consensus rates to the results of the IRF-based ABM varying the number of agents $N$ and arguments $M$.

%\newpage
\section{Conclusion}

% we enable compartmentalized coupling to disease or climate models

Our findings highlight the robustness of mean-field predictions and their compatibility with agent-based opinion models that explicitly incorporate cognitive structures (argument layer) and mechanisms (confirmation bias).
The results demonstrate that the mean-field approach accurately predicts the transition from consensus to polarization induced by increasing confirmation bias. This work advances our understanding of opinion dynamics in the presence of cognitive biases and offers a framework that can be adapted to study other biases and their impacts on social systems. In particular, the framework is applicable to a wide class of non-linear continuous opinion models \cite{Lorenz2021individual} that have been notoriously difficult to address with analytical tools.

By employing techniques from statistical physics, particularly mean-field theory, we provide a tractable and encompassing analysis of complex socio-cognitive dynamics. This approach, extensively reviewed in \cite{castellano2009statistical}, has been instrumental in modeling various social phenomena, including cultural dissemination \cite{castellano2000nonequilibrium}, the spread of innovations and diseases \cite{dodds2004universal,gleeson2013binary}.  
Future research can build on these findings to (i) explore more nuanced cognitive models, (ii) to build efficient compartment models which couple social dynamics to climate or disease models \cite{reitenbach2024coupled}, and (iii) to develop interventions aimed at mitigating the polarizing effects of cognitive biases in society.

%Non-linear models of opinion dynamics are notoriously difficult to address with analytical tools due to the intricate dependencies and feedback mechanisms inherent in such systems. However, our study demonstrates that by employing a mean-field treatment and reducing the complexity through compartmentalization, we can gain significant insights into the underlying dynamics. The agreement between our mean-field predictions and agent-based simulations underscores the efficacy of this approach. We have shown that despite the simplifications, the mean-field model captures the essential features of opinion dynamics, including critical points and phase transitions. This work not only provides a clearer understanding of how confirmation bias influences opinion evolution but also offers a framework that can be adapted to study other cognitive biases and their impacts on social systems. Future research can build on these findings to explore more nuanced models and to develop interventions aimed at mitigating the polarizing effects of cognitive biases in society.

%TC:break main
%the command above serves to have a word count for the abstract

%\begin{keywords}
%opinion dynamics | model synthesis
%\end{keywords}

%\begin{corrauthor}
%sven.banisch \at universecity.de
%\end{corrauthor}

\footnotesize
\textbf{Acknowledgement.}
We acknowledge funding from the German Federal Ministry for Education and Research (BMBF) for the infoXpand project (031L0300A,031L0300C), within the MONID consortium. 
The authors also acknowledge support by the state of Baden-Württemberg through the bwHPC.

%\section*{References}
%\bibliographystyle{apalike}
\bibliographystyle{apsrev4-1}
\bibliography{references.bib}

\begin{thebibliography}{34}%
\makeatletter
\providecommand \@ifxundefined [1]{%
 \@ifx{#1\undefined}
}%
\providecommand \@ifnum [1]{%
 \ifnum #1\expandafter \@firstoftwo
 \else \expandafter \@secondoftwo
 \fi
}%
\providecommand \@ifx [1]{%
 \ifx #1\expandafter \@firstoftwo
 \else \expandafter \@secondoftwo
 \fi
}%
\providecommand \natexlab [1]{#1}%
\providecommand \enquote  [1]{``#1''}%
\providecommand \bibnamefont  [1]{#1}%
\providecommand \bibfnamefont [1]{#1}%
\providecommand \citenamefont [1]{#1}%
\providecommand \href@noop [0]{\@secondoftwo}%
\providecommand \href [0]{\begingroup \@sanitize@url \@href}%
\providecommand \@href[1]{\@@startlink{#1}\@@href}%
\providecommand \@@href[1]{\endgroup#1\@@endlink}%
\providecommand \@sanitize@url [0]{\catcode `\\12\catcode `\$12\catcode `\&12\catcode `\#12\catcode `\^12\catcode `\_12\catcode `\%12\relax}%
\providecommand \@@startlink[1]{}%
\providecommand \@@endlink[0]{}%
\providecommand \url  [0]{\begingroup\@sanitize@url \@url }%
\providecommand \@url [1]{\endgroup\@href {#1}{\urlprefix }}%
\providecommand \urlprefix  [0]{URL }%
\providecommand \Eprint [0]{\href }%
\providecommand \doibase [0]{http://dx.doi.org/}%
\providecommand \selectlanguage [0]{\@gobble}%
\providecommand \bibinfo  [0]{\@secondoftwo}%
\providecommand \bibfield  [0]{\@secondoftwo}%
\providecommand \translation [1]{[#1]}%
\providecommand \BibitemOpen [0]{}%
\providecommand \bibitemStop [0]{}%
\providecommand \bibitemNoStop [0]{.\EOS\space}%
\providecommand \EOS [0]{\spacefactor3000\relax}%
\providecommand \BibitemShut  [1]{\csname bibitem#1\endcsname}%
\let\auto@bib@innerbib\@empty
%</preamble>
\bibitem [{\citenamefont {Axelrod}(1997)}]{axelrod1997dissemination}%
  \BibitemOpen
  \bibfield  {author} {\bibinfo {author} {\bibfnamefont {R.}~\bibnamefont {Axelrod}},\ }\href@noop {} {\bibfield  {journal} {\bibinfo  {journal} {Journal of conflict resolution}\ }\textbf {\bibinfo {volume} {41}},\ \bibinfo {pages} {203} (\bibinfo {year} {1997})}\BibitemShut {NoStop}%
\bibitem [{\citenamefont {Deffuant}\ \emph {et~al.}(2000)\citenamefont {Deffuant}, \citenamefont {Neau}, \citenamefont {Amblard},\ and\ \citenamefont {Weisbuch}}]{deffuant2000mixing}%
  \BibitemOpen
  \bibfield  {author} {\bibinfo {author} {\bibfnamefont {G.}~\bibnamefont {Deffuant}}, \bibinfo {author} {\bibfnamefont {D.}~\bibnamefont {Neau}}, \bibinfo {author} {\bibfnamefont {F.}~\bibnamefont {Amblard}}, \ and\ \bibinfo {author} {\bibfnamefont {G.}~\bibnamefont {Weisbuch}},\ }\href@noop {} {\bibfield  {journal} {\bibinfo  {journal} {Advances in Complex Systems}\ }\textbf {\bibinfo {volume} {3}},\ \bibinfo {pages} {87} (\bibinfo {year} {2000})}\BibitemShut {NoStop}%
\bibitem [{\citenamefont {Hegselmann}\ and\ \citenamefont {Krause}(2002)}]{hegselmann2002opinion}%
  \BibitemOpen
  \bibfield  {author} {\bibinfo {author} {\bibfnamefont {R.}~\bibnamefont {Hegselmann}}\ and\ \bibinfo {author} {\bibfnamefont {U.}~\bibnamefont {Krause}},\ }\href@noop {} {\bibfield  {journal} {\bibinfo  {journal} {Journal of Artificial Societies and Social Simulation}\ }\textbf {\bibinfo {volume} {5}},\ \bibinfo {pages} {1} (\bibinfo {year} {2002})}\BibitemShut {NoStop}%
\bibitem [{\citenamefont {Flache}\ \emph {et~al.}(2017)\citenamefont {Flache}, \citenamefont {M\"{a}s}, \citenamefont {Feliciani}, \citenamefont {Chattoe-Brown}, \citenamefont {Deffuant}, \citenamefont {Huet},\ and\ \citenamefont {Lorenz}}]{Flache2017models}%
  \BibitemOpen
  \bibfield  {author} {\bibinfo {author} {\bibfnamefont {A.}~\bibnamefont {Flache}}, \bibinfo {author} {\bibfnamefont {M.}~\bibnamefont {M\"{a}s}}, \bibinfo {author} {\bibfnamefont {T.}~\bibnamefont {Feliciani}}, \bibinfo {author} {\bibfnamefont {E.}~\bibnamefont {Chattoe-Brown}}, \bibinfo {author} {\bibfnamefont {G.}~\bibnamefont {Deffuant}}, \bibinfo {author} {\bibfnamefont {S.}~\bibnamefont {Huet}}, \ and\ \bibinfo {author} {\bibfnamefont {J.}~\bibnamefont {Lorenz}},\ }\href {\doibase 10.18564/jasss.3521} {\bibfield  {journal} {\bibinfo  {journal} {Journal of Artificial Societies and Social Simulation}\ }\textbf {\bibinfo {volume} {20}},\ \bibinfo {pages} {2} (\bibinfo {year} {2017})}\BibitemShut {NoStop}%
\bibitem [{\citenamefont {Banisch}\ and\ \citenamefont {Olbrich}(2019)}]{Banisch2019opinion}%
  \BibitemOpen
  \bibfield  {author} {\bibinfo {author} {\bibfnamefont {S.}~\bibnamefont {Banisch}}\ and\ \bibinfo {author} {\bibfnamefont {E.}~\bibnamefont {Olbrich}},\ }\href@noop {} {\bibfield  {journal} {\bibinfo  {journal} {The Journal of Mathematical Sociology}\ }\textbf {\bibinfo {volume} {43}},\ \bibinfo {pages} {76} (\bibinfo {year} {2019})}\BibitemShut {NoStop}%
\bibitem [{\citenamefont {Banisch}\ and\ \citenamefont {Shamon}(0)}]{Banisch2023biased}%
  \BibitemOpen
  \bibfield  {author} {\bibinfo {author} {\bibfnamefont {S.}~\bibnamefont {Banisch}}\ and\ \bibinfo {author} {\bibfnamefont {H.}~\bibnamefont {Shamon}},\ }\href@noop {} {\bibfield  {journal} {\bibinfo  {journal} {Sociological Methods \& Research}\ }\textbf {\bibinfo {volume} {0}},\ \bibinfo {pages} {00491241231186658} (\bibinfo {year} {0})}\BibitemShut {NoStop}%
\bibitem [{\citenamefont {Lord}\ \emph {et~al.}(1979)\citenamefont {Lord}, \citenamefont {Ross},\ and\ \citenamefont {Lepper}}]{lord1979biased}%
  \BibitemOpen
  \bibfield  {author} {\bibinfo {author} {\bibfnamefont {C.~G.}\ \bibnamefont {Lord}}, \bibinfo {author} {\bibfnamefont {L.}~\bibnamefont {Ross}}, \ and\ \bibinfo {author} {\bibfnamefont {M.~R.}\ \bibnamefont {Lepper}},\ }\href@noop {} {\bibfield  {journal} {\bibinfo  {journal} {Journal of personality and social psychology}\ }\textbf {\bibinfo {volume} {37}},\ \bibinfo {pages} {2098} (\bibinfo {year} {1979})}\BibitemShut {NoStop}%
\bibitem [{\citenamefont {Nickerson}(1998)}]{nickerson1998confirmation}%
  \BibitemOpen
  \bibfield  {author} {\bibinfo {author} {\bibfnamefont {R.~S.}\ \bibnamefont {Nickerson}},\ }\href@noop {} {\bibfield  {journal} {\bibinfo  {journal} {Review of general psychology}\ }\textbf {\bibinfo {volume} {2}},\ \bibinfo {pages} {175} (\bibinfo {year} {1998})}\BibitemShut {NoStop}%
\bibitem [{\citenamefont {Knobloch-Westerwick}\ \emph {et~al.}(2015)\citenamefont {Knobloch-Westerwick}, \citenamefont {Mothes}, \citenamefont {Johnson}, \citenamefont {Westerwick},\ and\ \citenamefont {Donsbach}}]{knobloch2015political}%
  \BibitemOpen
  \bibfield  {author} {\bibinfo {author} {\bibfnamefont {S.}~\bibnamefont {Knobloch-Westerwick}}, \bibinfo {author} {\bibfnamefont {C.}~\bibnamefont {Mothes}}, \bibinfo {author} {\bibfnamefont {B.~K.}\ \bibnamefont {Johnson}}, \bibinfo {author} {\bibfnamefont {A.}~\bibnamefont {Westerwick}}, \ and\ \bibinfo {author} {\bibfnamefont {W.}~\bibnamefont {Donsbach}},\ }\href@noop {} {\bibfield  {journal} {\bibinfo  {journal} {Journal of Communication}\ }\textbf {\bibinfo {volume} {65}},\ \bibinfo {pages} {489} (\bibinfo {year} {2015})}\BibitemShut {NoStop}%
\bibitem [{\citenamefont {Ling}(2020)}]{ling2020confirmation}%
  \BibitemOpen
  \bibfield  {author} {\bibinfo {author} {\bibfnamefont {R.}~\bibnamefont {Ling}},\ }\href@noop {} {\bibfield  {journal} {\bibinfo  {journal} {Digital Journalism}\ }\textbf {\bibinfo {volume} {8}},\ \bibinfo {pages} {596} (\bibinfo {year} {2020})}\BibitemShut {NoStop}%
\bibitem [{\citenamefont {Dandekar}\ \emph {et~al.}(2013)\citenamefont {Dandekar}, \citenamefont {Goel},\ and\ \citenamefont {Lee}}]{Dandekar2013biased}%
  \BibitemOpen
  \bibfield  {author} {\bibinfo {author} {\bibfnamefont {P.}~\bibnamefont {Dandekar}}, \bibinfo {author} {\bibfnamefont {A.}~\bibnamefont {Goel}}, \ and\ \bibinfo {author} {\bibfnamefont {D.~T.}\ \bibnamefont {Lee}},\ }\href@noop {} {\bibfield  {journal} {\bibinfo  {journal} {Proceedings of the National Academy of Sciences}\ }\textbf {\bibinfo {volume} {110}},\ \bibinfo {pages} {5791} (\bibinfo {year} {2013})}\BibitemShut {NoStop}%
\bibitem [{\citenamefont {Lorenz}\ \emph {et~al.}(2021)\citenamefont {Lorenz}, \citenamefont {Neumann},\ and\ \citenamefont {Schr{\"o}der}}]{Lorenz2021individual}%
  \BibitemOpen
  \bibfield  {author} {\bibinfo {author} {\bibfnamefont {J.}~\bibnamefont {Lorenz}}, \bibinfo {author} {\bibfnamefont {M.}~\bibnamefont {Neumann}}, \ and\ \bibinfo {author} {\bibfnamefont {T.}~\bibnamefont {Schr{\"o}der}},\ }\href@noop {} {\bibfield  {journal} {\bibinfo  {journal} {Psychological Review}\ } (\bibinfo {year} {2021})}\BibitemShut {NoStop}%
\bibitem [{\citenamefont {Lefebvre}\ \emph {et~al.}(2024)\citenamefont {Lefebvre}, \citenamefont {Deroy},\ and\ \citenamefont {Bahrami}}]{lefebvre2024roots}%
  \BibitemOpen
  \bibfield  {author} {\bibinfo {author} {\bibfnamefont {G.}~\bibnamefont {Lefebvre}}, \bibinfo {author} {\bibfnamefont {O.}~\bibnamefont {Deroy}}, \ and\ \bibinfo {author} {\bibfnamefont {B.}~\bibnamefont {Bahrami}},\ }\href@noop {} {\bibfield  {journal} {\bibinfo  {journal} {Proceedings of the Royal Society B}\ }\textbf {\bibinfo {volume} {291}},\ \bibinfo {pages} {20232011} (\bibinfo {year} {2024})}\BibitemShut {NoStop}%
\bibitem [{\citenamefont {Kermack}\ and\ \citenamefont {McKendrick}(1927)}]{kermack1927contribution}%
  \BibitemOpen
  \bibfield  {author} {\bibinfo {author} {\bibfnamefont {W.~O.}\ \bibnamefont {Kermack}}\ and\ \bibinfo {author} {\bibfnamefont {A.~G.}\ \bibnamefont {McKendrick}},\ }\href@noop {} {\bibfield  {journal} {\bibinfo  {journal} {Proceedings of the royal society of london. Series A, Containing papers of a mathematical and physical character}\ }\textbf {\bibinfo {volume} {115}},\ \bibinfo {pages} {700} (\bibinfo {year} {1927})}\BibitemShut {NoStop}%
\bibitem [{\citenamefont {Amit}\ \emph {et~al.}(1985)\citenamefont {Amit}, \citenamefont {Gutfreund},\ and\ \citenamefont {Sompolinsky}}]{amit1985spin}%
  \BibitemOpen
  \bibfield  {author} {\bibinfo {author} {\bibfnamefont {D.~J.}\ \bibnamefont {Amit}}, \bibinfo {author} {\bibfnamefont {H.}~\bibnamefont {Gutfreund}}, \ and\ \bibinfo {author} {\bibfnamefont {H.}~\bibnamefont {Sompolinsky}},\ }\href@noop {} {\bibfield  {journal} {\bibinfo  {journal} {Physical Review A}\ }\textbf {\bibinfo {volume} {32}},\ \bibinfo {pages} {1007} (\bibinfo {year} {1985})}\BibitemShut {NoStop}%
\bibitem [{\citenamefont {Castellano}\ \emph {et~al.}(2009)\citenamefont {Castellano}, \citenamefont {Fortunato},\ and\ \citenamefont {Loreto}}]{castellano2009statistical}%
  \BibitemOpen
  \bibfield  {author} {\bibinfo {author} {\bibfnamefont {C.}~\bibnamefont {Castellano}}, \bibinfo {author} {\bibfnamefont {S.}~\bibnamefont {Fortunato}}, \ and\ \bibinfo {author} {\bibfnamefont {V.}~\bibnamefont {Loreto}},\ }\href@noop {} {\bibfield  {journal} {\bibinfo  {journal} {Reviews of modern physics}\ }\textbf {\bibinfo {volume} {81}},\ \bibinfo {pages} {591} (\bibinfo {year} {2009})}\BibitemShut {NoStop}%
\bibitem [{\citenamefont {M{\"a}s}\ and\ \citenamefont {Flache}(2013)}]{Maes2013differentiation}%
  \BibitemOpen
  \bibfield  {author} {\bibinfo {author} {\bibfnamefont {M.}~\bibnamefont {M{\"a}s}}\ and\ \bibinfo {author} {\bibfnamefont {A.}~\bibnamefont {Flache}},\ }\href@noop {} {\bibfield  {journal} {\bibinfo  {journal} {PloS one}\ }\textbf {\bibinfo {volume} {8}},\ \bibinfo {pages} {e74516} (\bibinfo {year} {2013})}\BibitemShut {NoStop}%
\bibitem [{\citenamefont {Banisch}\ and\ \citenamefont {Olbrich}(2021)}]{Banisch2021argument}%
  \BibitemOpen
  \bibfield  {author} {\bibinfo {author} {\bibfnamefont {S.}~\bibnamefont {Banisch}}\ and\ \bibinfo {author} {\bibfnamefont {E.}~\bibnamefont {Olbrich}},\ }\href@noop {} {\bibfield  {journal} {\bibinfo  {journal} {Journal of Artificial Societies and Social Simulation}\ }\textbf {\bibinfo {volume} {24}} (\bibinfo {year} {2021})}\BibitemShut {NoStop}%
\bibitem [{\citenamefont {Taillandier}\ \emph {et~al.}(2021)\citenamefont {Taillandier}, \citenamefont {Salliou},\ and\ \citenamefont {Thomopoulos}}]{Taillandier2021introducing}%
  \BibitemOpen
  \bibfield  {author} {\bibinfo {author} {\bibfnamefont {P.}~\bibnamefont {Taillandier}}, \bibinfo {author} {\bibfnamefont {N.}~\bibnamefont {Salliou}}, \ and\ \bibinfo {author} {\bibfnamefont {R.}~\bibnamefont {Thomopoulos}},\ }\href@noop {} {\bibfield  {journal} {\bibinfo  {journal} {Journal of Artificial Societies and Social Simulation}\ }\textbf {\bibinfo {volume} {24}} (\bibinfo {year} {2021})}\BibitemShut {NoStop}%
\bibitem [{\citenamefont {Lorenz}(2007)}]{Lorenz2007continuous}%
  \BibitemOpen
  \bibfield  {author} {\bibinfo {author} {\bibfnamefont {J.}~\bibnamefont {Lorenz}},\ }\href@noop {} {\bibfield  {journal} {\bibinfo  {journal} {International Journal of Modern Physics C}\ }\textbf {\bibinfo {volume} {18}},\ \bibinfo {pages} {1819} (\bibinfo {year} {2007})}\BibitemShut {NoStop}%
\bibitem [{\citenamefont {Lopez-Pintado}\ and\ \citenamefont {Watts}(2008)}]{lopez2008social}%
  \BibitemOpen
  \bibfield  {author} {\bibinfo {author} {\bibfnamefont {D.}~\bibnamefont {Lopez-Pintado}}\ and\ \bibinfo {author} {\bibfnamefont {D.~J.}\ \bibnamefont {Watts}},\ }\href@noop {} {\bibfield  {journal} {\bibinfo  {journal} {Rationality and Society}\ }\textbf {\bibinfo {volume} {20}},\ \bibinfo {pages} {399} (\bibinfo {year} {2008})}\BibitemShut {NoStop}%
\bibitem [{\citenamefont {Reitenbach}\ \emph {et~al.}(2024)\citenamefont {Reitenbach}, \citenamefont {Sartori}, \citenamefont {Banisch}, \citenamefont {Golovin}, \citenamefont {Valdez}, \citenamefont {Kretzschmar}, \citenamefont {Priesemann},\ and\ \citenamefont {M{\"a}s}}]{reitenbach2024coupled}%
  \BibitemOpen
  \bibfield  {author} {\bibinfo {author} {\bibfnamefont {A.}~\bibnamefont {Reitenbach}}, \bibinfo {author} {\bibfnamefont {F.}~\bibnamefont {Sartori}}, \bibinfo {author} {\bibfnamefont {S.}~\bibnamefont {Banisch}}, \bibinfo {author} {\bibfnamefont {A.}~\bibnamefont {Golovin}}, \bibinfo {author} {\bibfnamefont {A.~C.}\ \bibnamefont {Valdez}}, \bibinfo {author} {\bibfnamefont {M.}~\bibnamefont {Kretzschmar}}, \bibinfo {author} {\bibfnamefont {V.}~\bibnamefont {Priesemann}}, \ and\ \bibinfo {author} {\bibfnamefont {M.}~\bibnamefont {M{\"a}s}},\ }\href@noop {} {\bibfield  {journal} {\bibinfo  {journal} {Reports on Progress in Physics}\ } (\bibinfo {year} {2024})},\ \bibinfo {note} {in press}\BibitemShut {NoStop}%
\bibitem [{\citenamefont {Burnstein}\ and\ \citenamefont {Vinokur}(1977)}]{burnstein1977persuasive}%
  \BibitemOpen
  \bibfield  {author} {\bibinfo {author} {\bibfnamefont {E.}~\bibnamefont {Burnstein}}\ and\ \bibinfo {author} {\bibfnamefont {A.}~\bibnamefont {Vinokur}},\ }\href@noop {} {\bibfield  {journal} {\bibinfo  {journal} {Journal of experimental social psychology}\ }\textbf {\bibinfo {volume} {13}},\ \bibinfo {pages} {315} (\bibinfo {year} {1977})}\BibitemShut {NoStop}%
\bibitem [{\citenamefont {Hinsz}\ and\ \citenamefont {Davis}(1984)}]{hinsz1984persuasive}%
  \BibitemOpen
  \bibfield  {author} {\bibinfo {author} {\bibfnamefont {V.~B.}\ \bibnamefont {Hinsz}}\ and\ \bibinfo {author} {\bibfnamefont {J.~H.}\ \bibnamefont {Davis}},\ }\href@noop {} {\bibfield  {journal} {\bibinfo  {journal} {Personality and Social Psychology Bulletin}\ }\textbf {\bibinfo {volume} {10}},\ \bibinfo {pages} {260} (\bibinfo {year} {1984})}\BibitemShut {NoStop}%
\bibitem [{\citenamefont {Sunstein}(2002)}]{sunstein2002law}%
  \BibitemOpen
  \bibfield  {author} {\bibinfo {author} {\bibfnamefont {C.~R.}\ \bibnamefont {Sunstein}},\ }\href@noop {} {\bibfield  {journal} {\bibinfo  {journal} {Journal of Political Philosophy}\ }\textbf {\bibinfo {volume} {10}} (\bibinfo {year} {2002})}\BibitemShut {NoStop}%
\bibitem [{\citenamefont {Feliciani}\ \emph {et~al.}(2020)\citenamefont {Feliciani}, \citenamefont {Flache},\ and\ \citenamefont {M{\"a}s}}]{Feliciani2020persuasion}%
  \BibitemOpen
  \bibfield  {author} {\bibinfo {author} {\bibfnamefont {T.}~\bibnamefont {Feliciani}}, \bibinfo {author} {\bibfnamefont {A.}~\bibnamefont {Flache}}, \ and\ \bibinfo {author} {\bibfnamefont {M.}~\bibnamefont {M{\"a}s}},\ }\href@noop {} {\bibfield  {journal} {\bibinfo  {journal} {Computational and Mathematical Organization Theory}\ ,\ \bibinfo {pages} {1}} (\bibinfo {year} {2020})}\BibitemShut {NoStop}%
\bibitem [{\citenamefont {Banisch}\ and\ \citenamefont {Shamon}(2024)}]{banisch2024validating}%
  \BibitemOpen
  \bibfield  {author} {\bibinfo {author} {\bibfnamefont {S.}~\bibnamefont {Banisch}}\ and\ \bibinfo {author} {\bibfnamefont {H.}~\bibnamefont {Shamon}},\ }\href@noop {} {\bibfield  {journal} {\bibinfo  {journal} {Journal of Artificial Societies \& Social Simulation}\ }\textbf {\bibinfo {volume} {27}} (\bibinfo {year} {2024})}\BibitemShut {NoStop}%
\bibitem [{Note1()}]{Note1}%
  \BibitemOpen
  \bibinfo {note} {Note that this rescaling has to be carried over to Equation (\ref {eq:argadoption}) by adjusting the parameter that governs the strength of biased processing $\beta \rightarrow M \beta $. Hence, while $\beta = 0.5$ has been identified as a critical value in a model with $M = 4$ pro and con arguments \protect \citep {Banisch2023biased}, this corresponds to $\beta = 2$ after rescaling.}\BibitemShut {Stop}%
\bibitem [{Note2()}]{Note2}%
  \BibitemOpen
  \bibinfo {note} {Note that, in principle, the probability of leaving a bi-polarized state should still be non-zero??}\BibitemShut {Stop}%
\bibitem [{\citenamefont {Castellano}\ \emph {et~al.}(2000)\citenamefont {Castellano}, \citenamefont {Marsili},\ and\ \citenamefont {Vespignani}}]{castellano2000nonequilibrium}%
  \BibitemOpen
  \bibfield  {author} {\bibinfo {author} {\bibfnamefont {C.}~\bibnamefont {Castellano}}, \bibinfo {author} {\bibfnamefont {M.}~\bibnamefont {Marsili}}, \ and\ \bibinfo {author} {\bibfnamefont {A.}~\bibnamefont {Vespignani}},\ }\href@noop {} {\bibfield  {journal} {\bibinfo  {journal} {Physical Review Letters}\ }\textbf {\bibinfo {volume} {85}},\ \bibinfo {pages} {3536} (\bibinfo {year} {2000})}\BibitemShut {NoStop}%
\bibitem [{\citenamefont {Dodds}\ and\ \citenamefont {Watts}(2004)}]{dodds2004universal}%
  \BibitemOpen
  \bibfield  {author} {\bibinfo {author} {\bibfnamefont {P.~S.}\ \bibnamefont {Dodds}}\ and\ \bibinfo {author} {\bibfnamefont {D.~J.}\ \bibnamefont {Watts}},\ }\href@noop {} {\bibfield  {journal} {\bibinfo  {journal} {Physical review letters}\ }\textbf {\bibinfo {volume} {92}},\ \bibinfo {pages} {218701} (\bibinfo {year} {2004})}\BibitemShut {NoStop}%
\bibitem [{\citenamefont {Gleeson}(2013)}]{gleeson2013binary}%
  \BibitemOpen
  \bibfield  {author} {\bibinfo {author} {\bibfnamefont {J.~P.}\ \bibnamefont {Gleeson}},\ }\href@noop {} {\bibfield  {journal} {\bibinfo  {journal} {Physical Review X}\ }\textbf {\bibinfo {volume} {3}},\ \bibinfo {pages} {021004} (\bibinfo {year} {2013})}\BibitemShut {NoStop}%
\bibitem [{\citenamefont {Friedkin}\ and\ \citenamefont {Johnsen}(2011)}]{Friedkin2011social}%
  \BibitemOpen
  \bibfield  {author} {\bibinfo {author} {\bibfnamefont {N.~E.}\ \bibnamefont {Friedkin}}\ and\ \bibinfo {author} {\bibfnamefont {E.~C.}\ \bibnamefont {Johnsen}},\ }\href@noop {} {\emph {\bibinfo {title} {Social influence network theory: A sociological examination of small group dynamics}}},\ Vol.~\bibinfo {volume} {33}\ (\bibinfo  {publisher} {Cambridge University Press},\ \bibinfo {year} {2011})\BibitemShut {NoStop}%
\bibitem [{Note3()}]{Note3}%
  \BibitemOpen
  \bibinfo {note} {We have confirmed this choices with numerical integration of the respective MF models.}\BibitemShut {Stop}%
\end{thebibliography}%

\clearpage
%\onecolumn
\appendix

\section{Reducing the PAT model to the IRF model}
\label{sec:SI:A}

\subsection{Naive approach}
\label{sec:SI:A:naive}

Let us denote the probability that the sender $j$ chooses an argument that supports a positive or negative stance by $P[+|o_j]$ and $P[-|o_j]$ respectively.
These probabilities depend on the opinion of $j$ and are given by
\begin{equation}
    \begin{array}{ll}
        P[+|o_j] = & \frac{1 + o_j}{2} \vspace{6pt}\\
        P[-|o_j] = & \frac{1 - o_j}{2}.
    \end{array}
    %P[+|o_j] = \frac{M - m o_j}{2 M}
    %P[m|o_j] = \frac{M - m o_j}{2 M}
\end{equation}
Notice again that disbelief $b_{jk} = 0$ in a con argument is considered to support a positive opinion.
For this reason the probabilities $P[+|o_j]$ and $P[-|o_j]$ are equal for all belief strings that map onto the same opinion.

Next, we have to compute the probability that an articulated belief $b_{jk}$ provides new information for agent $i$ (i.e. $b_{jk} \neq b_{ik}$).
This is more complicated as the probability depends on how many beliefs the two agents already share. In other words, this probability is not reducible to opinions $o_i$ and $o_j$, because it may differ for different pairs of belief strings $b_i$ and $b_j$ that realize the same opinions.
As a naive approximation, however, we shall assume no explicit belief dependence and write 
\begin{equation}
    \begin{array}{ll}
        P[new|+,o_i] = & \frac{1 - o_i}{2} \vspace{6pt}\\
        P[new|-,o_i] = & \frac{1 + o_i}{2}
    \end{array}
    %P[new|m,o_i] = \frac{M + m o_i}{2 M}
    %P[new|m,o_i] = \frac{M + m o_i}{2 M}
\end{equation}
for arguments that point into the positive or negative direction respectively.
Implicitly, this approximation entails the assumption that the exact position ($k$) of the beliefs does not play a role, and that the opinion of an agent is defined by an unordered set of pro and con arguments.

Opinion change takes place if the information sent by $j$ is new to and accepted by the receiver $i$.
In that case agent $i$ will change opinion by $1/M$ or respectively $-1/M$.
The probability for these two opinion change events can hence be computed as
\begin{equation}
    \begin{array}{ll}
        P[\Delta o_i = +\frac{1}{M} | o_i,o_j ] &= P[+|o_j] \ P[new|+,o_i] \  p_{\beta}(o_i,+)
        \vspace{6pt}\\
        P[\Delta o_i = -\frac{1}{M} | o_i,o_j ] &= P[-|o_j] \ P[new|-,o_i] \ p_{\beta}(o_i,-),
    \end{array}
\end{equation}
where the acceptance probabilities is given as before by $p_{\beta}(m,o_i)$ as specified in Equation (\ref{eq:argadoption}) with $m = 1$ for positive contributions and $m = -1$ for negative ones.
The expected opinion change is then given by 
\begin{equation}
    \begin{array}{ll}
        \mathrm{E}[\Delta o_i|o_i,o_j]  = f(o_i,o_j) & \vspace{6pt}\\
        = \frac{1}{M} \Big( P[\Delta o_i = +\frac{1}{M} | o_i,o_j ] - P[\Delta o_i = -\frac{1}{M} | o_i,o_j ] \Big). & \vspace{6pt}
    \end{array}
\end{equation}
Using $p_{\beta}(o_i,+) - p_{\beta}(o_i,-) = \tanh(\frac{o_i \beta}{2})$, we obtain
\begin{equation}
    \mathrm{E}[\Delta o_i|o_i,o_j] = \frac{1}{4M} \Big[ o_j - o_i + \tanh\Big(\frac{o_i \beta}{2}\Big) \Big(1 - o_i o_j\Big)\Big].
    \label{eq:SI:naivesolution}
\end{equation}

Notice that for $\beta = 0$ (no confirmation bias), Eq. \ref{eq:SI:naivesolution} reduces to
\begin{equation}
    \mathrm{E}[\Delta o_i|o_i,o_j] = \frac{1}{4M}(o_j - o_i)
\end{equation}
which corresponds to linear averaging with $\mu = 1/4M$ widely used in continuous opinion dynamics \citep{Friedkin2011social,deffuant2000mixing,Flache2017models}.

\subsection{Sophisticated approach}
\label{sec:SI:A:sophi}

Let's again assume that two agents $i$ and $j$ have been chosen.
As noted above, the probability that the sender $j$ supplies the receiver $i$ with \emph{new} information is non-trivial in the sense that the exact assignment of beliefs $b_i$ and $b_j$ does matter. 

%This is illustrated in Figure \ref{??}.
%\todo[inline]{Introduce distribution plots from Joris here.}

We first consider the case that $j$ supplies $i$ with a new argument in favor of the issue. As noted above, favorable information may correspond to a new pro argument ($b_{jk} = 1, b_{ik} = 0$ for $1 \leq k \leq M$) or to a counter argument that is dismissed ($b_{jk} = 0, b_{ik} = 1$ for $M+1 \leq k \leq 2M$). Hence, the probability of new positive information is
\begin{equation}
    P[ j \xrightarrow[]{+} i] =
    \frac{1}{2M} \Big[ \sum_{k=1}^M (1-b_{ik}) b_{jk}
    + \sum_{k=M+1}^{2M} (1-b_{jk}) b_{ik}
    \Big]
    \label{eq:SI:newpositive01}
\end{equation}
For further convenience, we introduce the bracket notation $\langle \rangle_p$ and $\langle \rangle_c$ as the normalized sum over the pro and the con part of the belief strings
\begin{align}
    \langle b_i \rangle_p &= \frac{1}{M}\sum_{k=1}^M b_{ik} \\
    \langle b_i \rangle_c &= \frac{1}{M}\sum_{k=M+1}^{2M} b_{ik}.
\end{align}
Besides simplifying the description, these operators have a direct interpretation as $o_i = \langle b_i \rangle_p - \langle b_i \rangle_c$. 
With this convention, Eq. (\ref{eq:SI:newpositive01}) can be written as
\begin{equation}
    P[ j \xrightarrow[]{+} i] =
    \frac{1}{2} \Big( \langle b_j \rangle_p - \langle b_i b_j \rangle_p + \langle b_i \rangle_c - \langle b_i b_j \rangle_c
    \Big),
    \label{eq:SI:newpositive02}
\end{equation}
where the two terms $\langle b_i b_j \rangle_p$ and $\langle b_i b_j \rangle_c$ are the normalized sums over element-wise products of $i$'s and $j$'s beliefs. Hence, these operators correspond to the normalized number of shared pro and con arguments respectively.

Equivalently, we can compute the probability that $j$ supplies $i$ with new information supporting a negative opinion by
\begin{equation}
    P[ j \xrightarrow[]{-} i] =
    \frac{1}{2M} \Big[ \sum_{k=1}^M (1-b_{jk}) b_{ik}
    + \sum_{k=M+1}^{2M} (1-b_{ik}) b_{jk}
    \Big],
    \label{eq:SI:newnegative01}
\end{equation}
which can be written as
\begin{equation}
    P[ j \xrightarrow[]{-} i] =
    \frac{1}{2} \Big( \langle b_i \rangle_p - \langle b_i b_j \rangle_p + \langle b_j \rangle_c - \langle b_i b_j \rangle_c
    \Big).
    \label{eq:SI:newnegative02}
\end{equation}

With $P[ j \xrightarrow[]{+} i]$ and $P[ j \xrightarrow[]{-} i]$ it is easy to compute the expected opinion change. As in the naive setting, $o_i$ will increase by $1/M$ if the agent accepts a new positive information, and decrease by this amount when accepting negative information
\begin{equation}
    \begin{array}{ll}
        P[\Delta o_i = +\frac{1}{M} | o_i,o_j ] &= P[ j \xrightarrow[]{+} i] \  p_{\beta}(o_i,+)
        \vspace{6pt}\\
        P[\Delta o_i = -\frac{1}{M} | o_i,o_j ] &= P[ j \xrightarrow[]{-} i] \ p_{\beta}(o_i,-).
    \end{array}
\end{equation}
The expected opinion change is then given by
\begin{equation}
    \mathrm{E}[\Delta o_i|o_i,o_j] = \frac{p_{\beta}(o_i,+) P[ j \xrightarrow[]{+} i] - p_{\beta}(o_i,-) P[ j \xrightarrow[]{-} i]}{M}
    \label{eq:SI:sophisticatedsolution}
\end{equation}

\paragraph{No confirmation bias.} For $\beta = 0$, we have $p_{\beta}(o_i,+) = p_{\beta}(o_i,-) = 1/2$ and hence all higher order terms from Eq. \eqref{eq:SI:newpositive02} and \eqref{eq:SI:newnegative02} cancel. We obtain
\begin{equation}
    \mathrm{E}[\Delta o_i|o_i,o_j] = \frac{1}{4 M}\Big(
    \langle b_j \rangle_p - \langle b_j \rangle_c + \langle b_i \rangle_c - \langle b_i \rangle_p\Big).
    \label{eq:SI:nobias01}
\end{equation}
With $o_i = \langle b_i \rangle_p - \langle b_i \rangle_c$ and $o_j = \langle b_j \rangle_p - \langle b_j \rangle_c$ we recover the naive solution
\begin{equation}
    \mathrm{E}[\Delta o_i|o_i,o_j] = \frac{1}{4M}(o_j - o_i)
    \label{eq:SI:nobias02}
\end{equation}

\paragraph{With confirmation bias.} With $\beta > 0$ the situation is slightly more complex because the higher-order terms $\langle b_i b_j\rangle_p$ and $\langle b_i b_j\rangle_c$ do not cancel. For further formal transformation we make use of the covariance relations $\sigma_p(b_i,b_j) = \langle b_i b_j\rangle_p - \langle b_i\rangle_p \langle b_j\rangle_p$ and $\sigma_c(b_i,b_j) = \langle b_i b_j\rangle_c - \langle b_i\rangle_c \langle b_j\rangle_c$. With that, we arrive at 
\begin{equation}
    \mathrm{E}[\Delta o_i|b_i,b_j] = \frac{1}{4M} \Big[ o_j - o_i + \tanh\Big(\frac{o_i \beta}{2}\Big) \Big(1 - o_i o_j + \hat{y}(b_i,b_j)\Big)\Big] 
\end{equation}
where 
\begin{equation}
\begin{aligned}
\hat{y}(b_i,b_j)
    &= 1 - \langle b_i\rangle_p - \langle b_i\rangle_c - \langle b_j\rangle_p - \langle b_j\rangle_c \\ & + \langle b_i\rangle_p \langle b_j\rangle_p + \langle b_i\rangle_c \langle b_j\rangle_c \\  & + \langle b_i\rangle_p \langle b_j\rangle_c + \langle b_i\rangle_c \langle b_j\rangle_p 
    \\  & + 2 (\sigma_p(b_i,b_j) + \sigma_c(b_i,b_j))       
\end{aligned}
\end{equation}
is a correction of the naive solution in \eqref{eq:SI:naivesolution} that still depends on the argument strings $b_i$ and $b_j$ and is not reducible to opinions.
Hence, the term $\hat{y}(b_i,b_j)$ informs about the error that is made by the approximate naive solution, and we will analyze this error in the next section.

\subsection{Analytical and numerical interrogation}
\label{sec:SI:A:inter}

%\todo[inline]{Joris, how many combination are there? There are $2^{2M} * 2^{2M}$ possible belief combinations.}

To prove that the naive approximation provides a valid IRF for the PAT model in the sense that it captures its mean behavior, we have to show that the mean of $\hat{y}(b_i,b_j)$ over all belief strings $b_i$ and $b_j$ that map onto the same opinions $o_i$ and $o_j$ is zero for any opinion pair. For any pair of belief strings, $\hat{y}(b_i,b_j)$ captures the difference between the naive and sophisticated solution (\eqref{eq:SI:naivesolution} and \eqref{eq:SI:sophisticatedsolution} respectively). For the case with $M = 4$, we explicitly calculated $\hat{y}(b_i,b_j)$ for every possible belief combination ($2^{2M} \times 2^{2M} = 2^{16}$ possibilities). For any pair of opinions $o_i$ and $o_j$, we find 
\begin{equation}
    \mathbb{E}_{o_i o_j}[ \hat{y} ] = \sum_{ (b_i,b_j) \rightarrow (o_i,o_j) } \hat{y}(b_i,b_j) = 0
\end{equation}
showing that the individual "errors" ($\hat{y}$), present for specific belief string combinations, average out due to the different correlations at the belief level.

\section{Impact of the number of arguments $M$}
\label{sec:B}

An important parameter present in both the IRF and PAT model version is the number of arguments (denoted with $M$). While the arguments are not explicitly modeled in the IRF model, $M$ still influences the expected opinion change in the denominator of the IRF function. Its influence on the presence of polarization among agents is studied in the following section. In particular, we show that (i.) the meta-stability of bi-polarization in the PAT model \cite{Banisch2023biased} vanishes if the number of arguments $M$ becomes larger, and (ii.) that these transient effects are not present in the reduced IRF-based model.

\begin{figure}[htbp]
    \centering
    \includegraphics[width=0.5\textwidth]{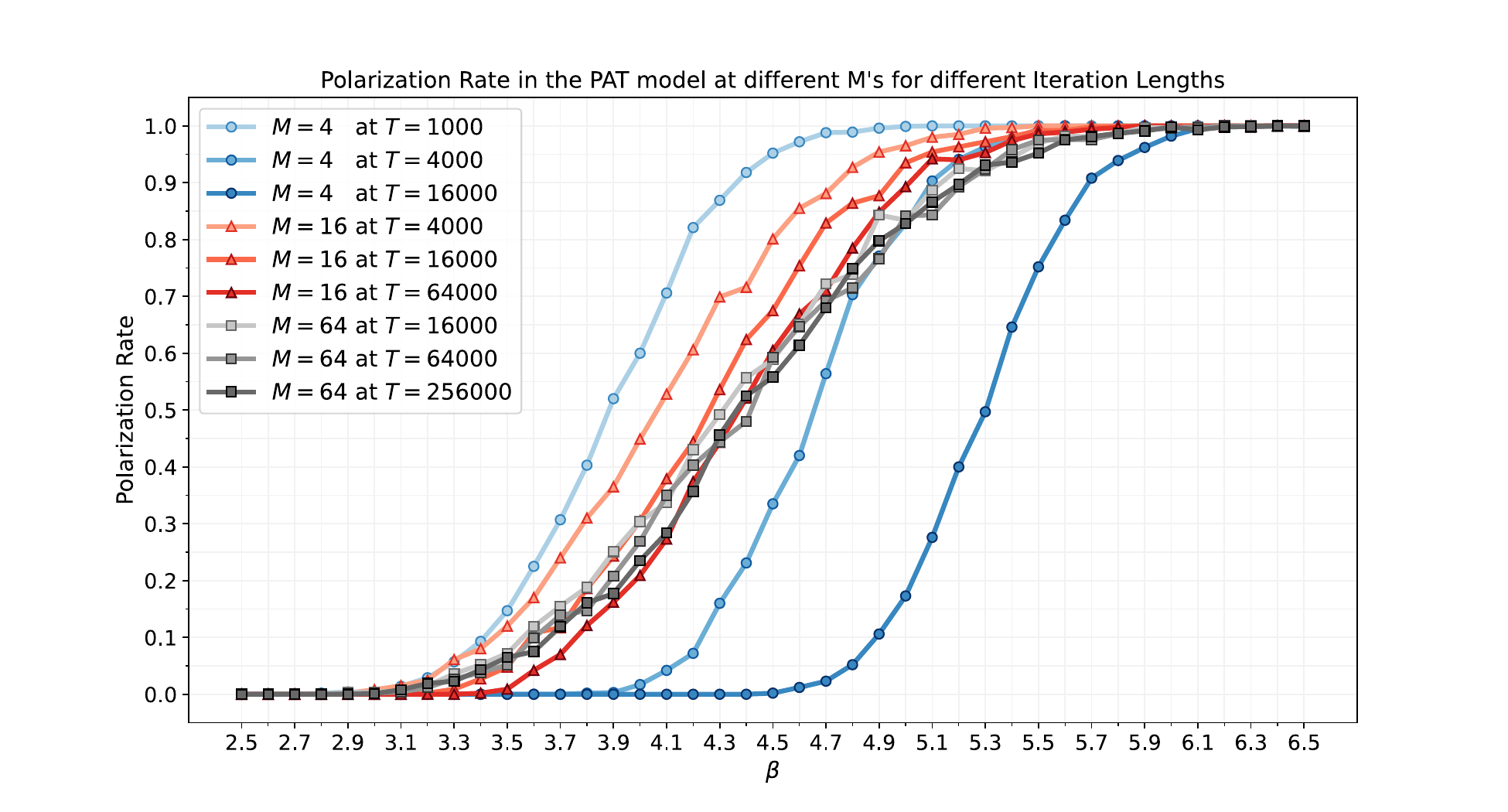}
    \caption{Average polarization rate for different $\beta$, $M$ and $T$ in the PAT model. We use 40 sample points for $\beta \in \{2.5, \dots, 6.5\}$ for each value of $M \in \{4,16,64\}$. For each combination of $\beta$ and $M$, the PAT model was run for $M \times T_{base} = T$ iterations where $T_{base} \in \{250, 1000, 4000\}$. Each parameter combination ($\beta$, $M$, $T$) was sampled 1000 times with $N=100$ and we show the average polarization rate over these samples.}
    \label{fig:M:pol_rate_ß_M_PAT}
\end{figure}

Figure~\ref{fig:M:pol_rate_ß_M_PAT} shows the average polarization rate in the PAT model based on 1000 samples for each combination ($\beta$, $M$, $T = T_{base} \times M$) where $\beta \in \{2.5,\ 2.6,\ \dots,\ 6.5\}$, $M \in \{4,\ 16,\ 64\}$ and $T_{base} \in \{250,\ 1000,\ 4000\}$. For each $M$, a separate color is chosen. The shade of a color differentiates between the different $T_{base}$. The total number of iterations is defined as $T = T_{base} \times M$. A different number of iterations $T$ is utilized for the different $M$ to account for its general influence on the size of expected opinion changes $1/(4 M)$, and hence time to reach an attractor.\footnote{We have confirmed this choices with numerical integration of the respective MF models.} For comparability between simulations with different $M$, $T_{base}$ is introduced.

For a small $M = 4$ (as used throughout the paper), we reproduce the findings from Fig. \ref{fig:panel_runtime} (l.h.s.) in the main text. These findings are consistent with previous work \cite{Banisch2023biased}, which has shown that the time a bi-polarized state persists increases exponentially with $\beta$. However, Fig. \ref{fig:M:pol_rate_ß_M_PAT} shows that this effect vanishes when the number of arguments increases to $M = 64$. The reason is that a small number of arguments leads to a considerable probability that agents on one side of the opinion spectrum subsequently adopt arguments from the other side in a series of random events.

Looking at the lowest $T_{base} = 250$ (leading to $T=1000,\ 4000,\ 16000$ for the different $M$s), it is possible to see, that increasing $M$ delays the consensus-polarization transition with respect to $\beta$.

\begin{figure}[htbp]
    \centering
    \includegraphics[width=0.5\textwidth]{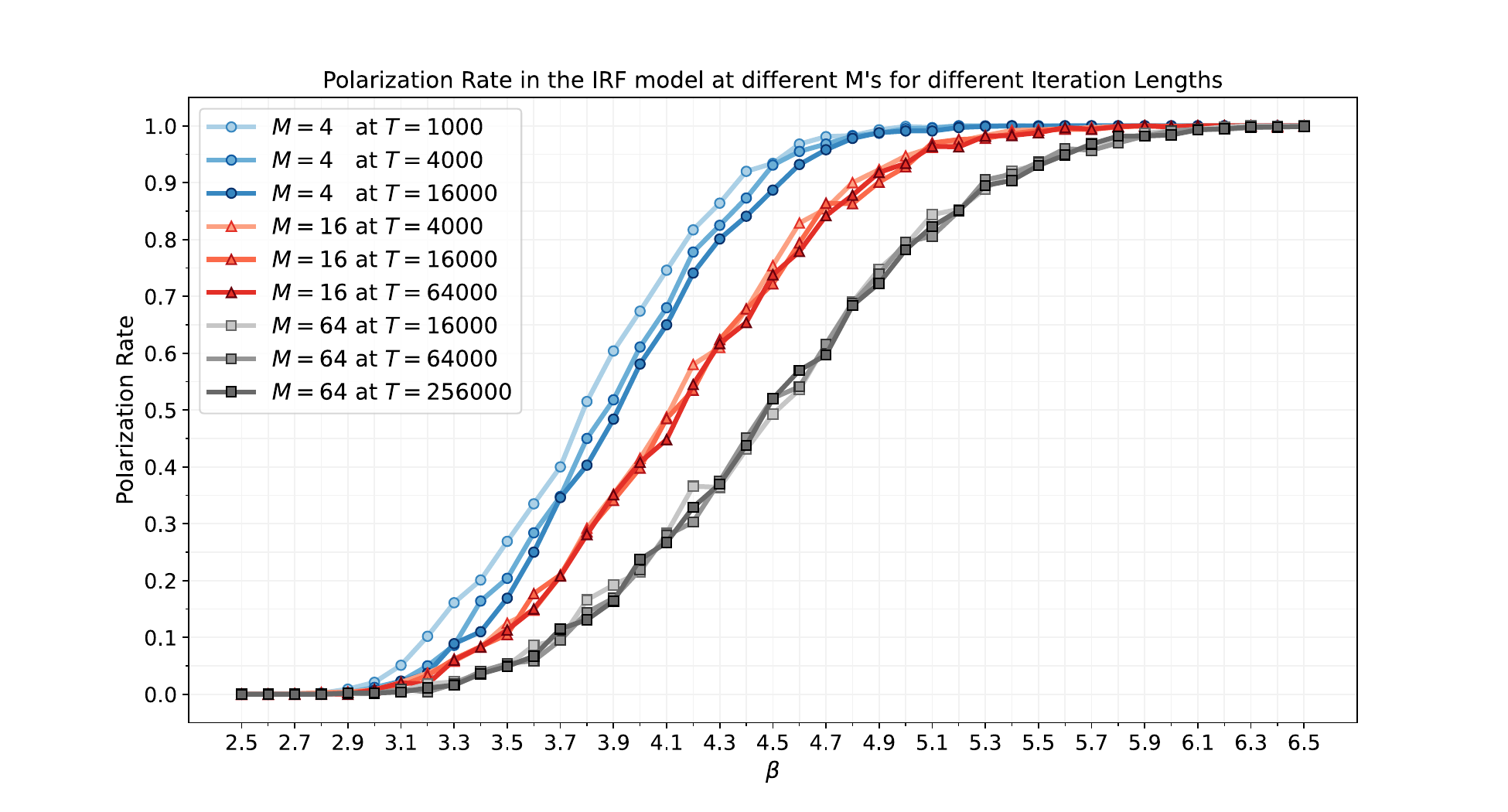}
    \caption{Average polarization rate for different $\beta$, $M$ and $T$ in the IRF model. We use 40 sample points for $\beta \in \{2.5, \dots, 6.5\}$ for each value of $M \in \{4,16,64\}$. For each combination of $\beta$ and $M$, the IRF model was run for $M \times T_{base} = T$ iterations where $T_{base} \in \{250, 1000, 4000\}$. Each parameter combination ($\beta$, $M$, $T$) was sampled 1000 times with $N=100$ and we show the average polarization rate over these samples.}
    \label{fig:M:pol_rate_ß_M_IRF}
\end{figure}

We follow the same procedure for the IRF model shown in Figure~\ref{fig:M:pol_rate_ß_M_IRF}. The same parameter space was sampled, again with 1000 samples per ($\beta$, $M$, $T$). For each $M$ a separate color is chosen and the shade of the color corresponds to the different $T_{base}$. We first note that an increase in $M$, on average, leads to a delayed polarization with respect to $\beta$. In the main text, we show that this delay of the consensus-polarization transition with respect to $\beta$ is captured by the proposed MF approach.
Secondly, increasing the iteration length $T$ for a given $M$ does not affect the polarization rate. This reproduces the findings from Fig. \ref{fig:panel_runtime} (r.h.s.) in the main text.

\begin{figure}[htbp]
    \centering
    \includegraphics[width=0.99\linewidth]{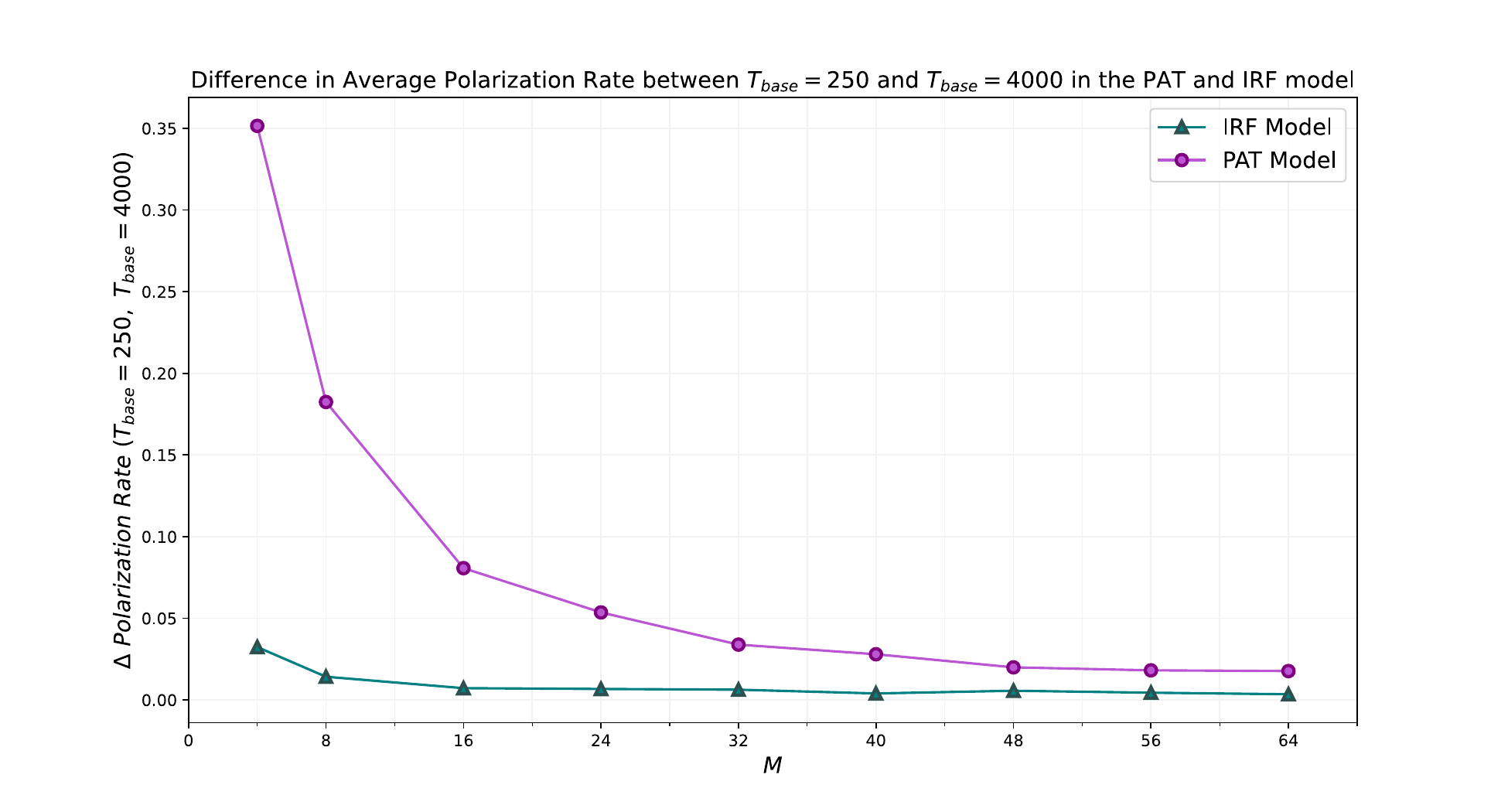}
    \caption{Difference in the average polarization rate for different $M$ in the PAT and IRF model. We run 1000 simulations per parameter point ($\beta$, $M$, $T = T_{base} * M$) with $\beta \in \{2.5,\ 2.6,\ \dots,\ 6.5\}$, $M \in \{4,\ 8,\ 16,\ 24,\ \dots,\ 64\}$ and $T_{base} \in \{250,\ 4000\}$ for the PAT model and the IRF model. The average polarization rate over the different samples and $\beta$s was calculated for each value of $M$ and $T_{base} = 250,\ 4000$. The plot shows the difference in the average polarization rate between the two $T_{base}$. This is the proportion of realizations that collapse from polarization at time $T = 250 \times M$ into extreme consensus at $T = 4000 \times M$}
    \label{fig:M:diff_pol_rate_M}
\end{figure}

Finally, Fig.~\ref{fig:M:diff_pol_rate_M} shows the difference in the polarization rate between $T_{base}=250$ and $T_{base}=4000$ for $M=4,\ 8,\ 16,\ 24,\ \dots,\ 64$. This measure hence assesses how many realizations collapse from a bi-polarized state into consensus between the two time points. For each $M$, the average difference between the polarization rates are evaluated over the sampled $\beta$s. Increasing $M$ diminishes the transient effects in the PAT model. Further inquiry has to clarify if the remaining average difference $\Delta P_{250}^{4000}$ converges to a non-zero baseline (as indicated by this analysis). In the IRF model, the differences are relatively small for all sampled $M$, underlining the stability of polarization in the IRF model regardless of the number of arguments. 

All in all, this section shows that the transient effects related to the meta-stability of the polarization attractor for $M = 4$, are rather a special case within the parameter space of the model.

\section{Computational Setting for Fig. \ref{fig:basicbifurcation}}

Fig. \ref{fig:basicbifurcation} shows the consensus-polarization transition comparing all three models. For the MF theory, it shows the bifurcation diagram in terms of the variance of the opinion profile ($o = (o_A,o_B)$) in the fixed points of the MF model. Stable fixed points are represented by solid curves, unstable points by dashed curves. Results have been obtained with Mathematica 12.3 (NSolveValues) for $\beta = 0.02, 0.04, \ldots , 6$ (300 sample points). The results for both ABMs are based on simulations with $N = 1000$ and $M = 4$. The models are run for $T = 4000$ steps. Confirmation bias is sampled as $\beta = 0.0, 0.1, \ldots , 6$ (60 sample points). Per point 100 simulation runs have been performed. The scatter plots show the variance over the emerging opinion profile $o = (o_1,o_2, \ldots,o_N)$ for all individual runs, taking the temporal average over the last 400 steps (measurement period).

%\todo[inline]{The script is available on the github repository.}

% Main

%In the original argument model, bi-polarization is a meta-stable state that persists for an exponentially longer time period with increasing $\beta$ \cite{Banisch2023biased}. In the reduced model bi-polarization becomes stable.  

\section{MF theory for the ABM in the transition regime}
\label{sec:SI:D}

One of the puzzles when comparing simulation results to the bifurcation curves is that in the ABM only polarization is observed at large $\beta$. In the MF model, the extreme consensus point remains a stable fixed point for all $\beta$, and this equilibrium is reached whenever $o_A$ and $o_B$ tend to the same side of the opinion spectrum. 

\subsection{Initial conditions}
One hypothesis on this phenomena is that heterogeneous initial conditions of the ABM project into a particular region of the MF space. The phenomena suggests that, as $\beta$ grows large, initial conditions project into the basin of attraction of a polarizing fixed point. We test this by generating a large number (10000 samples in all experiments reported here) of initial conditions for different system sizes ($N = 100,1000,10000$). We divide the agent population in two groups such that group A contains all agents with $o_i < 0$ and groups B all agents with $o_i > 0$ (\emph{splitting at zero}). When we look on how the group means project into MF space, we observe that they are scattered in a specific region of the opinion space. In Fig. \ref{fig:ic:approach}, we show this for $N = 100$ agents initialized (i.) with random arguments (binomial opinion distribution), and (ii.) uniformly at random ($o_i \in [-1,1]$). 

%\begin{figure}[htbp]
%    \centering
%    \includegraphics[width=0.99\linewidth]{figures/initialconditions/approach.pdf}
%    \caption{....}
%    \label{fig:ic:approach}
%\end{figure}

\subsection{Basin of attraction}
Along with the initial conditions projected into MF space, we compute the basin of attraction for a given $\beta$. We do that numerically by sampling regularly through the opinion space $(o_A,o_B)$ ($200 \times 200$ sample points). For each point, we iterate the MF model for $T = 1000$ steps and differentiate polarized states from extreme consensus states using the variance. The two histograms on the left in Fig. \ref{fig:ic:approach} show that realization have clearly settled in either of these states after 1000 steps. The green-shaded region in the main panal of Fig. \ref{fig:ic:approach} corresponds to the basin of attraction associated with bi-polarization for $\beta =3.9$.

\subsection{Predicting consensus rate in transition regime}
It turns out that this hypothesis provides a compelling explanation for the transition behavior of the finite size heterogeneous opinion ABM. The higher $\beta$, the larger the basin of attraction and the more initial conditions project within that region. This explains why the ABM only ends up in the bi-polarizing attractors once $\beta$ is large enough. 

\begin{figure}[htbp]
    \centering
    \includegraphics[width=0.79\linewidth]{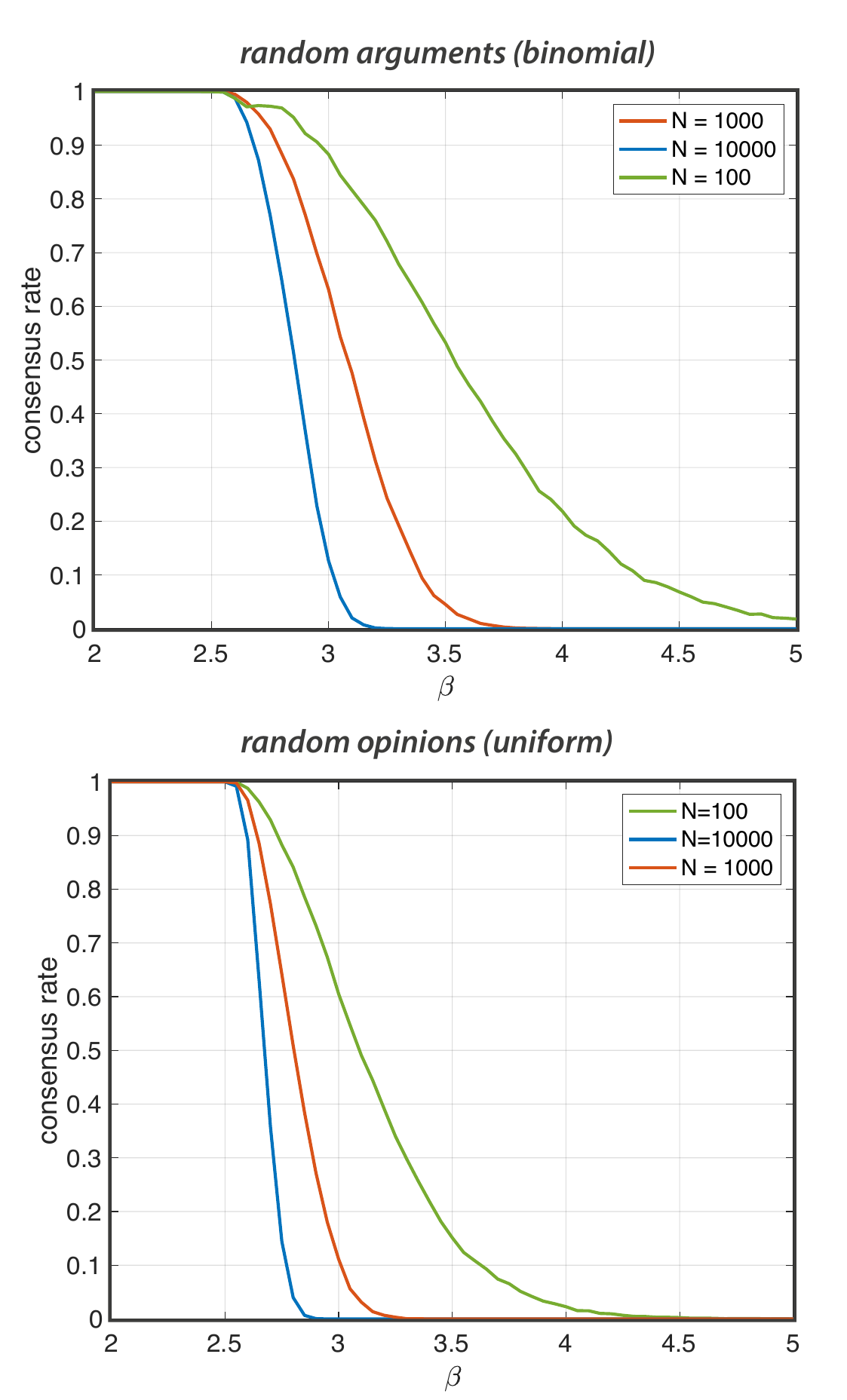}
    \caption{Mean field prediction of consensus rate for different initial conditions (random arguments versus random opinions) and system sizes ($N \in \{100,1000,10000\}$). }
    \label{fig:SI:ic:consensusratebeta}
\end{figure}

What is more, we can use that approach to predict the probability of bi-polarization (and inversely the consensus rate). As shown for $\beta = 3.9$ in Fig. \ref{fig:ic:approach}, the projected initial conditions cut across the basin of attraction associated to the fixed point at the upper left where $o_A \approx -1$ and $o_B \approx 1$. Under the MF assumptions, this means that a certain part of the initial conditions will evolve to consensus (blue dots), and another part to polarization (red dots). The respective distributions of final states are shown on the right of Fig. \ref{fig:ic:approach} for the two different initial conditions.
In order to obtain an estimate of consensus rates in transition regime based on MF theory we numerically sample through $\beta \in [2,5]$.
We use the proportion of initial conditions that end up in consensus versus as an indicator and study how it depends on the system size ($N$) and the confirmation bias ($\beta$).  
The results for $N = 100,1000,10000$ are shown in Fig. \ref{fig:SI:ic:consensusratebeta} for binomial and uniform initial conditions.

\begin{figure}[htbp]
    \centering
    \includegraphics[width=0.99\linewidth]{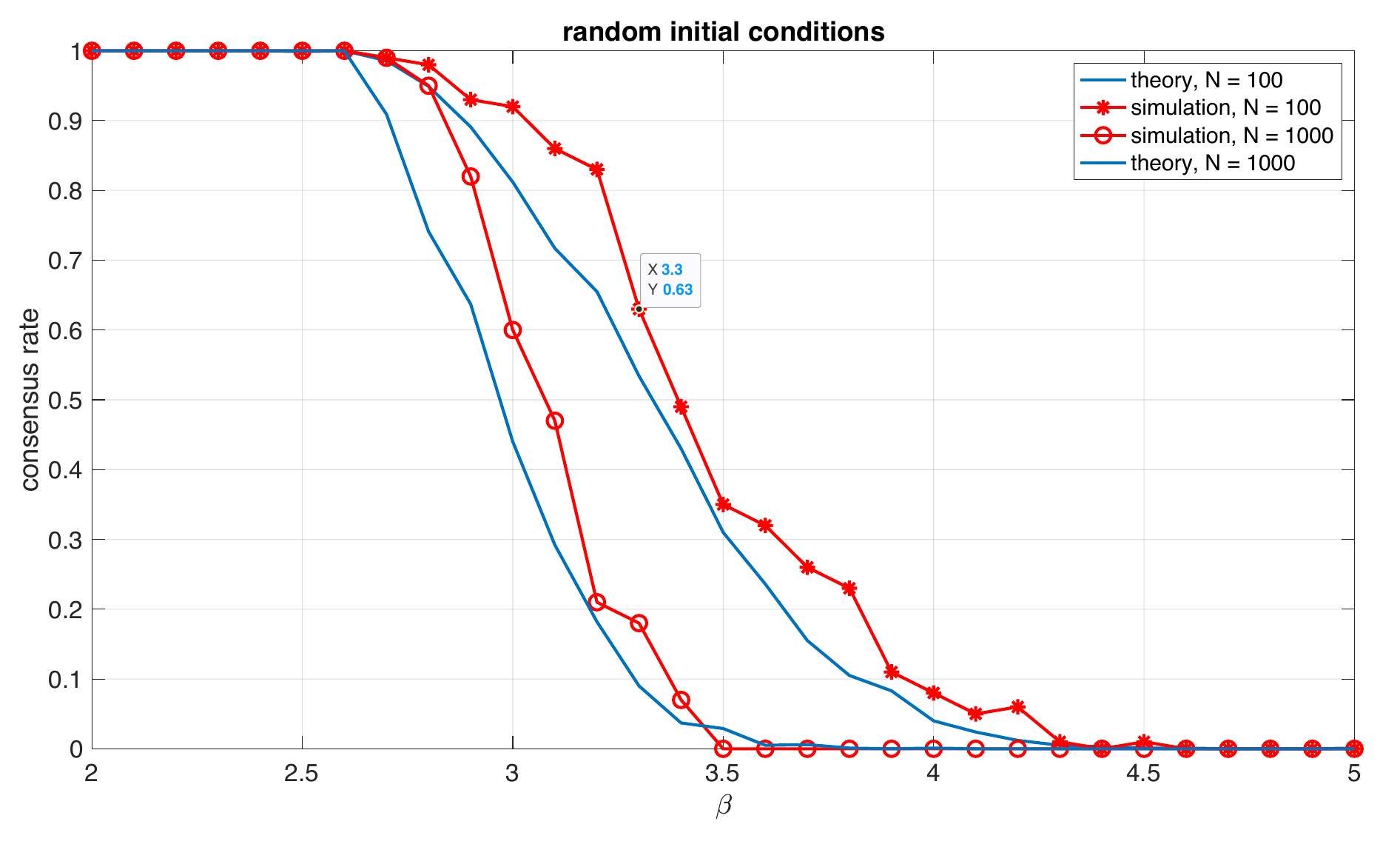}
    \caption{The MF approach accounting for group sizes works considerably well. Here we compare the MF prediction to simulations for random initial conditions.}
    \label{fig:SI:ic:gs_correction}
\end{figure}

%\begin{figure}[htbp]
%    \centering
%    \includegraphics[width=0.99\linewidth]{figures/initialconditions/SummaryCRDifferentSplitting.pdf}
%    \caption{....}
%    \label{fig:SI:ic:initialsumary}
%\end{figure}

%\subsection{Comparison to simulations}
%The theory makes specific predictions on how the ABM with heterogeneous initial conditions behaves in the bifurcation regime. But to what extend do these predictions match with the simulations?
%In order to test this we perform simulations of the ABM with the same conditions. The result is shown in Fig. \ref{}.

%We find the the theory underestimates the consensus rate in the transition regime such that fro a given $\beta$ more ABM simulation than expected reach a final extreme consensus equilibrium. Despite that quantitative mismatch, the analysis captures the fact that the transition becomes sharper with growing system size. It also captures systematic differences between binomial and random initial conditions.

\subsection{Group size correction}
In the previous test we constructed the two MF groups based on whether their opinion is positive or negative. (We have tested two alternative constructions based on the mean opinion and a half-to-half splitting of the agent population and encountered no significant differences.) But we have not yet taken into account that these two groups may differ in size. In order to do that, we have to use a mean field model in which group interaction probabilities take into account group sizes $s_A = N_A/N$ and $s_B = N_B/N$. For $p=1/2$ (random mixing), this is given by
\begin{equation}
    \Delta o_A = s_A f(o_A,o_A) + s_B f(o_A,o_B)
\end{equation}
and
\begin{equation}
    \Delta o_B = s_B f(o_B,o_B) + s_A f(o_B,o_A)
\end{equation}
We use the same procedure as before to predict consensus rates with the refined MF model.

%\bibliographystyle{apsrev4-1}
%\bibliography{references.bib}

\end{document}